\documentclass[12pt]{article}
\usepackage{cite}
\usepackage{hyperref}
\usepackage{amsmath,amssymb,amsfonts,amsthm,bm}
\usepackage{tabularx}
\usepackage{breqn}
\usepackage{graphicx}
\usepackage{floatrow}
\usepackage{lineno}
\usepackage[labelfont=bf]{caption}
\hypersetup{
    citecolor=blue,
    colorlinks=true,
    linkcolor=blue,
    filecolor=magenta,      
    urlcolor=blue,
}
\title{A hydrodynamic instability drives protein droplet formation on microtubules to nucleate branches}

\author
{\parbox{\linewidth}{\centering Sagar U. Setru$^{\dag, 1}$, Bernardo Gouveia$^{\dag, 2}$, Raymundo Alfaro-Aco$^3$, Joshua W. Shaevitz$^{\ast,1,4}$, Howard A. Stone$^{\ast, 5}$, Sabine Petry$^{\ast, 3}$}\\
\\
\normalsize{$^{1}$Lewis-Sigler Institute for Integrative Genomics}\\
\normalsize{$^{2}$Department of Chemical and Biological Engineering}\\
\normalsize{$^{3}$Department of Molecular Biology}\\
\normalsize{$^{4}$Department of Physics}\\
\normalsize{$^{5}$Department of Mechanical and Aerospace Engineering}\\
\normalsize{Princeton University, Princeton, NJ 08544, USA}\\
\\
\normalsize{$^{\dag}$These authors contributed equally.} \\
{\parbox{\linewidth}{\normalsize{$^{\ast}$To whom correspondence should be addressed; E-mail: \mbox{spetry@princeton.edu}, \mbox{hastone@princeton.edu}, \mbox{shaevitz@princeton.edu}.}}}}

\date{}

\begin{document}

\baselineskip24pt

\maketitle 

\textbf{Liquid-liquid phase separation\cite{Shin2017,alberti2019considerations} occurs not only in bulk liquid, but also on surfaces. In physiology, the nature and function of condensates on cellular structures remain unexplored. Here, we study how the condensed protein TPX2 behaves on microtubules to initiate branching microtubule nucleation\cite{Petry2013,Thawani2019,King2019}, which is critical for spindle assembly in eukaryotic cells\cite{uehara2010functional, uehara2009augmin, petry2011augmin,lawo2009haus, hotta2012characterization}. Using fluorescence, electron, and atomic force microscopies and hydrodynamic theory, we show that TPX2 on a microtubule reorganizes according to the Rayleigh-Plateau instability, like dew droplets patterning a spider web \cite{Rayleigh1878,Quere1999}. After uniformly coating microtubules, TPX2 forms regularly spaced droplets from which branches nucleate. Droplet spacing increases with greater TPX2 concentration. A stochastic model shows that droplets make branching nucleation more efficient by confining the space along the microtubule where multiple necessary factors colocalize to nucleate a branch.}

Branching microtubule nucleation plays a major role in spindle assembly and chromosome segregation in dividing eukaryotic cells, where it is required to generate microtubules in the spindle for kinetochore fiber tension, spindle bipolarity, and cytokinesis  \cite{uehara2010functional, uehara2009augmin, petry2011augmin,lawo2009haus, hotta2012characterization}. Its malfunction has been linked to a worse prognosis in cancer  \cite{perez2012mitotic,Uhleneaan2017}. The nucleation of a new microtubule from the side of a preexisting microtubule requires TPX2, the augmin complex, and the $\gamma$-tubulin ring complex ($\gamma$-TuRC) \cite{Petry2013}. The first component to bind to the preexisting microtubule is TPX2 \cite{Thawani2019}, which forms a liquid-like condensate on the microtubule; the condensate recruits tubulin and increases branching nucleation efficiency  \cite{King2019}. Other proteins also form condensed phases when associated with microtubules, such as Tau \cite{Hernandez-Vega2017, Siahaan2019} and BugZ \cite{Jiang2015}. Yet, how these proteins behave on the microtubule surface and how this behavior translates to biological function remain unexplored. Here, we investigate the dynamics of condensed TPX2 on the microtubule. We find that the hydrodynamic Rayleigh-Plateau instability causes TPX2 to form regularly spaced droplets along the microtubule. Then, microtubule branches nucleate from these droplets.

\begin{figure}
    \includegraphics[width=4.5in, height=5.5in]{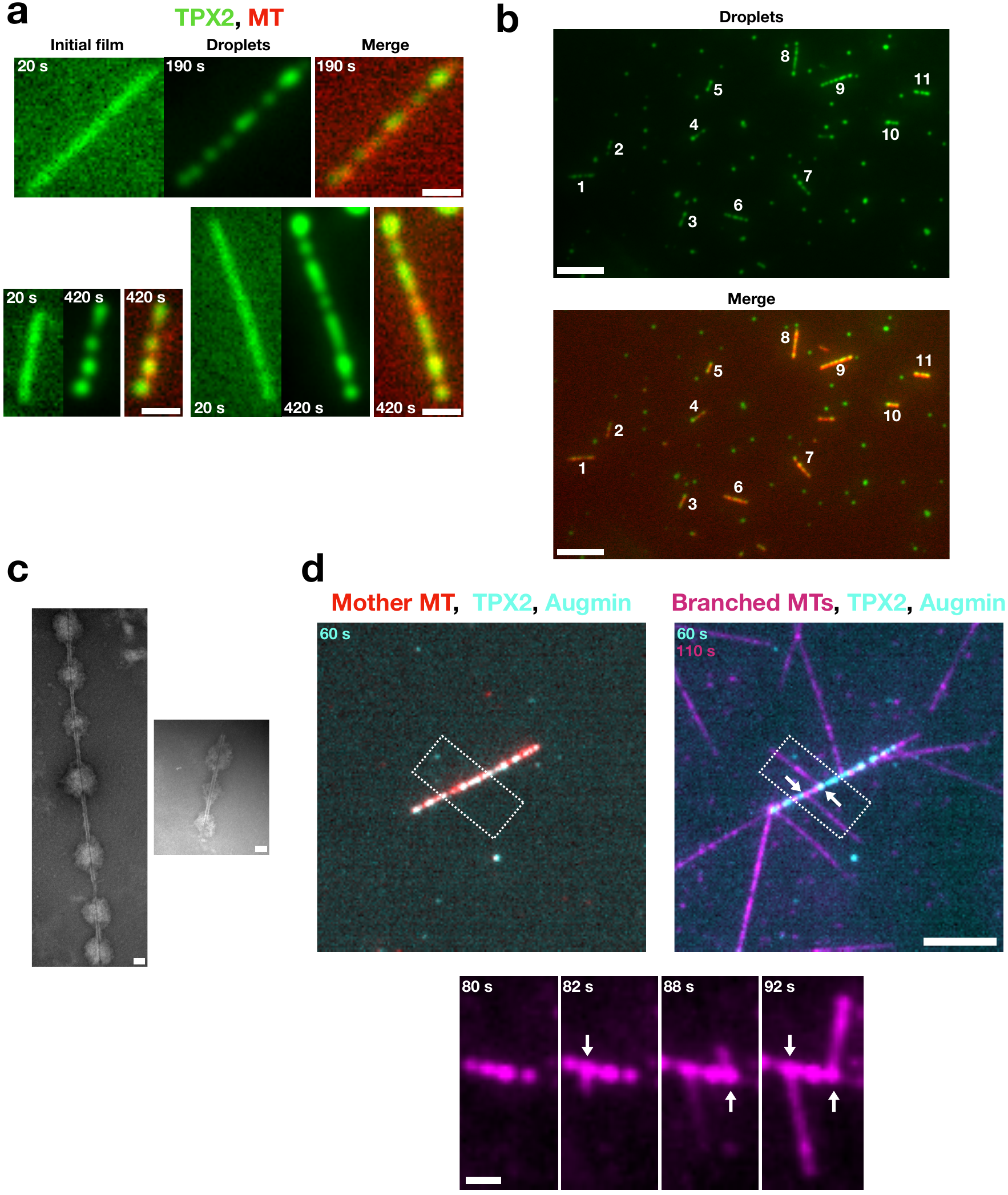}
  \caption{\textbf{TPX2 uniformly coats microtubules and then forms periodically spaced droplets that can nucleate branches.}
(a) Initial films and subsequent droplets of TPX2 on microtubules visualized using TIRF microscopy (\hyperref[movieCaptions]{Supplementary Movie 1}). $1~\mu\mathrm{M}$ GFP-TPX2 was spiked onto a passivated glass surface coated with Alexa568-labeled microtubules. Scale bars are $1~\mu\mathrm{m}$. (b)  Large field of view of a TIRF experiment after droplets have formed along microtubules. GFP-TPX2 concentration is 1 $\mu$M. Microtubules with a droplet pattern are marked with a number. Scale bars are 5 $\mu$m. (c) TPX2 droplets on microtubules imaged using electron microscopy. $0.1~\mu\mathrm{M}$ GFP-TPX2 was incubated with microtubules bound to a carbon grid. Scale bars are $100~\mathrm{nm}$. (d) Branched microtubules nucleating from TPX2 droplets formed along the initial mother microtubule, assembled \textit{in vitro} as in \cite{Alfaro-Aco2019} (\hyperref[movieCaptions]{Supplementary Movie 2}). $\gamma$-TuRC purified from  \textit{X. laevis} meiotic cytosol and recombinant GFP-augmin were included. Arrows indicate branched microtubules. Scale bars are $5~\mu\mathrm{m}$ (top) and $1~\mu\mathrm{m}$ (bottom). Only the soluble Cy5-tubulin channel (magenta) was imaged over time to enable a higher frame rate. The GFP-TPX2 and GFP-augmin channel (cyan) and the Alexa568 template microtubule channel (red) were only imaged at the start at $60$ s.}
 \label{fig:1}
\end{figure}

We first studied the dynamics of TPX2 binding to microtubules \textit{in vitro} using total internal reflection fluorescence (TIRF) microscopy (\hyperref[fig:1]{Fig.~1a}, \hyperref[fig:1]{Fig.~1b}, \hyperref[methods:tirf]{Methods}). GFP-TPX2 at a concentration of 1 $\mu$M formed an initially uniform coating along individual microtubules within seconds. This coating then broke up into a periodic pattern of droplets over tens of seconds with size $0.5 \pm 0.1~\mu\mathrm{m}$ and spacing $0.6 \pm 0.2~\mu\mathrm{m}$ (mean $\pm$ standard deviation, $N = 35$ microtubules) along individual microtubules (\hyperref[movieCaptions]{Supplementary Movie 1}, \hyperref[fig:s1]{Supplementary Fig.~1}). Similar patterns of condensed protein have also been previously observed on single microtubules for TPX2 \cite{King2019} and microtubule bundles for Tau \cite{Hernandez-Vega2017}, BugZ \cite{Jiang2015}, and LEM2 \cite{von2020lem2}. We next performed the same experiment at a lower, physiological concentration of TPX2, $0.1~\mu\mathrm{M}$ \cite{King2019,wuhr2014deep}. We observed a uniform coating but no visible droplet formation (\hyperref[fig:s2]{Supplementary Fig.~2a}). In contrast, at higher resolution, electron microscopy (\hyperref[methods:em]{Methods}) revealed that regularly spaced droplets do form at $0.1~\mu\mathrm{M}$, with size $0.29 \pm 0.03~\mu\mathrm{m}$ and spacing $0.46 \pm 0.11~\mu\mathrm{m}$ (mean $\pm$ standard deviation, $N = 2$ microtubules) (\hyperref[fig:1]{Fig.~1c}). This indicates that these droplets can exist below the diffraction limit of visible light. We then reconstituted branching microtubule nucleation \textit{in vitro} using purified proteins \cite{Alfaro-Aco2019} (\hyperref[methods:reconstitution]{Methods}) and observed that branches originate from TPX2 droplets colocalized with augmin and $\gamma$-TuRC (\hyperref[fig:1]{Fig.~1d}, \hyperref[movieCaptions]{Supplementary Movie 2}). $75 \pm 18\%$ of TPX2 droplets nucleated branches (mean $\pm$ standard deviation, $N = 7$ microtubules, \hyperref[table:s1]{Supplementary Table 1}).
Droplet formation always happened before the nucleation of a branch, and no branches nucleated from areas that did not have droplets.
Finally, in meiotic cytosol (\hyperref[methods:branching]{Methods}), microtubules also nucleate from a TPX2-coated microtubule to form a branched network (\hyperref[fig:s2]{Supplementary Fig.~2b}, \hyperref[movieCaptions]{Supplementary Movie 3}). These results suggest that droplet formation from condensed TPX2 may be important for branching microtubule nucleation.

\begin{figure}
\centering
\floatbox[{\capbeside\thisfloatsetup{capbesideposition={right,top},capbesidewidth=6cm}}]{figure}[\FBwidth]
{\caption{\textbf{AFM measurements reveal condensed TPX2 dynamics on microtubules.}
(a) Time-lapse AFM height topographies of TPX2 uniformly coating and then forming regularly spaced droplets on a microtubule. During data acquisition, GFP-TPX2 at $0.1\pm0.2~\mu\mathrm{M}$ (estimated range) was spiked onto microtubules adhered to a mica surface. Shown is the entire measured span of the microtubule in the top left of \hyperref[movieCaptions]{Supplementary Movie 4}. The topography was smoothed using a $40~\mathrm{nm} \times 40~\mathrm{nm}$ median filter. White carets mark droplets. Scale bar is $100~\mathrm{nm}$. (b) Height profiles centered on the microtubule long axis before coating (black), just after coating (blue), and when droplets have formed on the microtubule surface (red). Height profiles were smoothed using a moving-average window of $40~\mathrm{nm}$. The raw height profiles and their power spectra are shown in \hyperref[fig:s3]{Supplementary Fig.~3}. (c) Averaged power spectra calculated from the raw, unsmoothed height profiles across many microtubules for uncoated (black, $N=22$ microtubules), uniformly coated (blue, $N=23$ microtubules), and droplet-patterned microtubules (red, $N = 17$ microtubules). A peak is seen only in the data for droplet-patterned microtubules, corresponding to a droplet spacing of $260 \pm 20~\mathrm{nm}$ (mean $\pm$ standard deviation). Shaded regions represent $95\%$ bootstrap confidence intervals.}\label{fig:2}}
{\includegraphics[width=9cm]{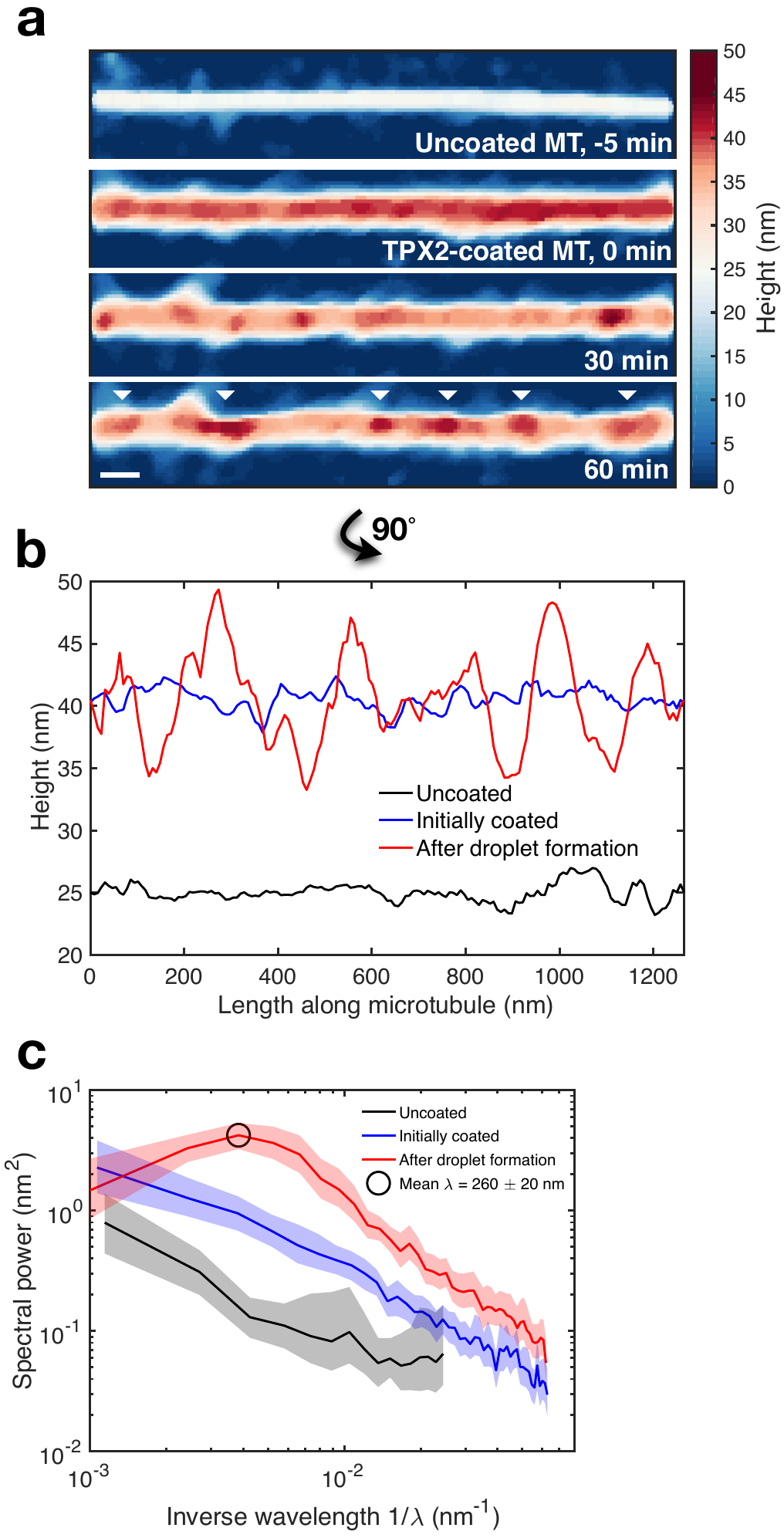}}
\end{figure}

We wished to study the dynamics of droplet formation of TPX2 alone on microtubules at higher spatial resolution than available by fluorescence microscopy and with temporal resolution not accessible by electron microscopy. Therefore, we turned to atomic force microscopy (AFM) to measure the topography of the initial coating and subsequent beading up of TPX2 on microtubules (\hyperref[methods:afm]{Methods}). By scanning the AFM tip over the sample every $\simeq4$ minutes, with a resolution of $\simeq8~\mathrm{nm}$ in the sample plane and $\simeq1~\mathrm{nm}$ in height, we measured the height of a bare microtubule on a mica surface to be $25 \pm 2~\mathrm{nm}$ (mean $\pm$ standard deviation) (\hyperref[fig:2]{Fig.~2a} $-5~\mathrm{min}$, \hyperref[fig:2]{Fig.~2b} black line), consistent with the known diameter of $25~\mathrm{nm}$ \cite{desai1997microtubule}. We then added TPX2 at a concentration of $0.2\pm0.1~\mu\mathrm{M}$ (estimated range). After addition of TPX2, the height signal uniformly increased to $41 \pm 3~\mathrm{nm}$ (mean $\pm$ standard deviation) as the condensed protein coated the microtubule (\hyperref[fig:2]{Fig.~2a} $0~\mathrm{min}$, \hyperref[fig:2]{Fig.~2b} blue line). The film of TPX2 then proceeded to bead up into a periodic pattern of droplets
along the microtubule with spacing $250 \pm 35~\mathrm{nm}$ (mean $\pm$ standard error of the mean) (\hyperref[fig:2]{Fig.~2a} $60~\mathrm{min}$, \hyperref[fig:2]{Fig.~2b} red line, \hyperref[movieCaptions]{Supplementary Movie 4}). The white carets in the $60~\mathrm{minute}$ topography in \hyperref[fig:2]{Fig.~2a} mark the droplets. The longer time scale to form droplets and the different spacings between droplets in AFM experiments compared to fluorescence and electron microscopy experiments is due to the different biochemical conditions and components used in each experimental method (\hyperref[table:s2]{Supplementary Table 2}).
The emergent periodicity of the condensate is evident in the power spectra of the raw height profile along the microtubule averaged over many samples (\hyperref[fig:2]{Fig.~2c}). Power spectra rely on the Fourier transform to identify the frequency components of a signal buried in noise (\hyperref[methods:afmDataAnalysis]{Methods}). Peaks in a power spectrum indicate the presence of a periodic pattern amidst noise; a monotonic power spectrum is expected for data that lacks periodicity. The power spectra (\hyperref[fig:2]{Fig.~2c}) show no characteristic length scale before and immediately after coating with TPX2, whereas a peak with wavelength $260 \pm 20~\mathrm{nm}$ (mean $\pm$ standard deviation) has emerged by $60$ minutes. Thus, the topography of condensed TPX2 on microtubules exhibits systematic emergent periodicity.

\begin{figure}
\centering
\floatbox[{\capbeside\thisfloatsetup{capbesideposition={right,top},capbesidewidth=6cm}}]{figure}[\FBwidth]
{\caption{\textbf{Hydrodynamic theory predicts TPX2 droplet formation on a microtubule surface.} 
(a) Schematic of the Rayleigh-Plateau instability. TPX2 initially coats the microtubule uniformly with thickness $h = r_{\mathrm{o}} - r_{\mathrm{i}}$. This film breaks up into droplets with spacing $\lambda_{\mathrm{max}}$ due to capillary forces on a time scale $r_{\mathrm{o}}\mu/\gamma$, where $\mu$ is the condensate viscosity and $\gamma$ is the surface tension. (b) Dispersion relation showing the growth rate versus wave number for different aspect ratios $r_{\mathrm{i}}/r_{\mathrm{o}}$. The most unstable mode (black circles and line) grows most quickly and corresponds to the observed droplet spacing $\lambda_{\mathrm{max}}$. (c) $\lambda_{\mathrm{max}}=f(h_{\mathrm{eff}})$ (solid orange line) and $\lambda_{\mathrm{max}}=f(h)$ (dotted orange line). Overlaid are AFM measurements of the film thickness and droplet spacing for many microtubules over a range of TPX2 concentrations. $\lambda_{\mathrm{max}}=f(h_{\mathrm{eff}})$ uses an effective height $h_{\mathrm{eff}}$, which is calculated to compensate for the reduced volume of condensate that coats the microtubule when the microtubule is resting on a surface. $\lambda_{\mathrm{max}}=f(h)$ is the exact prediction by theory. There are no fit parameters in either case. The shaded area encompasses the wavelengths that grow within 25\% of the maximum growth rate for each $h_{\mathrm{eff}}$. Error bars are standard error of the mean. For each concentration, we also plot the corresponding average power spectra (\hyperref[fig:s7]{Supplementary Fig.~7a}).}\label{fig:3}}
{\includegraphics[width=8cm]{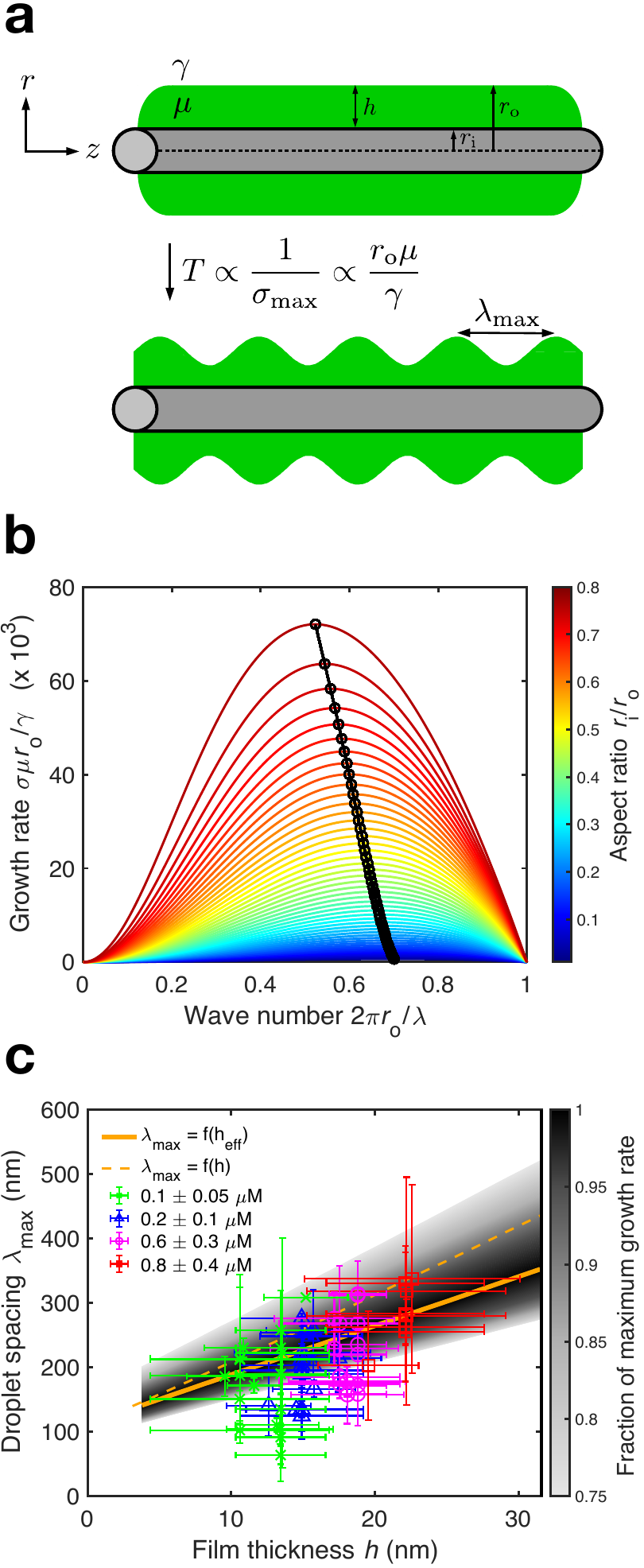}}
\end{figure}

Fluids that coat a solid fiber are known to form droplets via the Rayleigh-Plateau instability \cite{Rayleigh1878}. Surface tension causes the fluid film to be unstable due to the curvature of the filament surface and the surface area is minimized by forming periodically spaced droplets along the fiber \cite{Goren1962, Boulogne2012, Haefner2015, Quere1999}. Following Goren \cite{Goren1962}, but working directly at low Reynolds number as is appropriate for our experimental system, we solved a linear stability problem for the growth rate $\sigma$ of the droplet pattern as a function of the wave number $k = 2\pi/\lambda$, where $\lambda$ is the pattern wavelength (\hyperref[fig:3]{Fig.~3a}, \hyperref[fig:s4]{Supplementary Fig.~4}, \hyperref[theory:1-1]{Supplementary Theory}). We find that for a given ratio of the microtubule radius to the outer film radius, $r_{\mathrm{i}}/r_{\mathrm{o}}$, there is a wavelength $\lambda_{\mathrm{max}}$ that grows with the largest growth rate $\sigma_{\mathrm{max}}$ (\hyperref[fig:3]{Fig.~3b}). This wavelength will grow exponentially faster than all other wavelengths, leading to a periodic interface with wavelength $\lambda_{\mathrm{max}}$. Thus, we identify the thicker regions of the AFM height profiles as droplets formed by this instability.

We tested the ability of this theory to explain droplet formation on microtubules by measuring film thicknesses and subsequent droplet spacings at different bulk concentrations of TPX2 (\hyperref[fig:5]{Supplementary Fig.~5}). The radius of the microtubule is fixed at $r_{\mathrm{i}} = 25~\mathrm{nm}$. However, the thickness of the initial TPX2 film depends on its bulk concentration in solution and the density of microtubules (\hyperref[fig:s6]{Supplementary Fig.~6}, \hyperref[theory:1-2]{Supplementary Theory}). Capitalizing on this experimental fact, we changed the initial film thickness from $h = 13 \pm 2~\mathrm{nm}$ at $0.1\pm0.05~\mu\mathrm{M}$ TPX2 to $h = 22 \pm 1~\mathrm{nm}$ at $0.8\pm0.4~\mu\mathrm{M}$ TPX2 (mean $\pm$ standard deviation and estimated range, respectively) for a fixed microtubule density. The lower concentrations are physiological in healthy cells \cite{King2019,wuhr2014deep}. The higher concentrations may reflect overexpression in cancer tumor cells, in which TPX2 often has higher genetic copy number \cite{tonon2005high} and transcript and protein expression \cite{perez2012mitotic, neumayer2014tpx2}, and can be a negative prognostic indicator \cite{Uhleneaan2017}. TPX2 formed regularly spaced droplets along microtubules with consistently larger spacings $\lambda_{\mathrm{max}}$ as its bulk concentration increased, following theory (\hyperref[fig:3]{Fig.~3c}, \hyperref[fig:s7]{Supplementary Fig.~7a}, \hyperref[table:s3]{Supplementary Table 3}). 

The theory is purely geometric and has no free parameters. The predicted wavelength does not depend on the material properties of the TPX2 condensate such as viscosity or surface tension, which only set the timescale for pattern formation. We note that the higher microtubule density used in AFM experiments (\hyperref[fig:2]{Fig.~2}, \hyperref[fig:3]{Fig.~3c}) leads to thinner condensed films than in the EM and reconstitution experiments (\hyperref[fig:1]{Fig.~1}) and hence smaller droplet sizes, even at similar TPX2 concentrations (\hyperref[table:s2]{Supplementary Table 2}, \hyperref[theory:1-2]{Supplementary Theory}). We also note that our theory systematically predicts a larger wavelength than we measure (\hyperref[fig:3]{Fig.~3c}, dotted orange line). This is due to the adjacent surface underneath the microtubules during AFM imaging. The surface reduces the volume of fluid that can coat the microtubule; since $\lambda_{\mathrm{max}} \sim h \sim V^{1/3}$, a smaller volume produces a smaller wavelength.
Using trigonometry, we estimate this lost volume and plot the wavelength as a function of an effective height, $\lambda_{\mathrm{max}} = f(h_{\mathrm{eff}})$ (\hyperref[fig:3]{Fig.~3c}, solid orange line). We see that this effective curve compensates for the systematic offset that the data have with respect to the exact prediction by theory. 

Even with this correction, \hyperref[fig:3]{Fig.~3c} shows a large spread in the measured $\lambda_{\mathrm{max}}$. We identify two sources for the spread in our measurements. First, the dispersion relation we calculate (\hyperref[fig:3]{Fig.~3b}) has a broad peak, which means that wavelengths near the maximum growth rate $\lambda_{\mathrm{max}}$ will grow nearly as fast (\hyperref[fig:3]{Fig.~3c}, shaded area). Therefore, spread in the measured $\lambda_{\mathrm{max}}$ is a natural consequence of the hydrodynamic theory.
Second, low-force (25-40 pN), nanometric AFM in fluid is invariably susceptible to thermal noise. This is apparent in the raw height profiles (\hyperref[fig:s3]{Supplementary Fig.~3a}) and power spectra (\hyperref[fig:s3]{Supplementary Fig.~3b}) of the microtubule shown in \hyperref[fig:2]{Fig.~2a} and \hyperref[fig:2]{Fig.~2b}, as well as the power spectra of individual microtubules across TPX2 concentrations (\hyperref[fig:s5]{Supplementary Fig.~5}, right column).

In addition, we measured the growth rate of $\lambda_{\mathrm{max}}$ to be exponential at early times, as expected for a linear instability (\hyperref[fig:s7]{Supplementary Fig.~7b}). At later times, the periodicity has already been selected as the droplet pattern has set in. Thus, the spectral power versus time stops changing. We also see that the time to form droplets is orders of magnitude greater than the time to grow the initial film. The film grows more quickly because the timescale for its growth is set by fast diffusion of protein in the bulk, whereas the timescale for droplet formation is limited by the slow capillary velocity $\gamma/\mu$ of the condensate (\hyperref[theory:1-1]{Supplementary Theory}). As a control, kinesin-1, a motor protein that does not exist as a condensed phase in any known physiological context and whose binding site on the microtubule is structurally known \cite{cao2014structure, shang2014high}, did not exhibit hydrodynamic behavior on the microtubule as measured by AFM (\hyperref[fig:7]{Supplementary Fig.~7a}).

\begin{figure}
\centering
\floatbox[{\capbeside\thisfloatsetup{capbesideposition={right,top},capbesidewidth=6cm}}]{figure}[\FBwidth]
{\caption{\textbf{A stochastic model predicts that TPX2 droplets enhance the efficiency of branching microtubule nucleation.}
(a) Electron microscopy images show ring-shaped $\gamma$-TuRCs (black arrows) localizing to regularly spaced TPX2 droplets along the microtubule in the presence of augmin. Scale bars are $100~\mathrm{nm}$. (b) Schematic of branching factors binding to a microtubule coated with a uniform TPX2 layer versus periodic TPX2 droplets. (c) Monte Carlo simulations show that droplets colocalize two necessary factors faster than a uniform coating. Here, $k_{\mathrm{on}}l/k_{\mathrm{off}}$ = 10. These results are not sensitive to this parameter choice (\hyperref[fig:s9]{Supplementary Fig.~9}). Each datapoint is the average of $10^{7}$ independent simulations.}\label{fig:4}}
{\includegraphics[width=9.5cm]{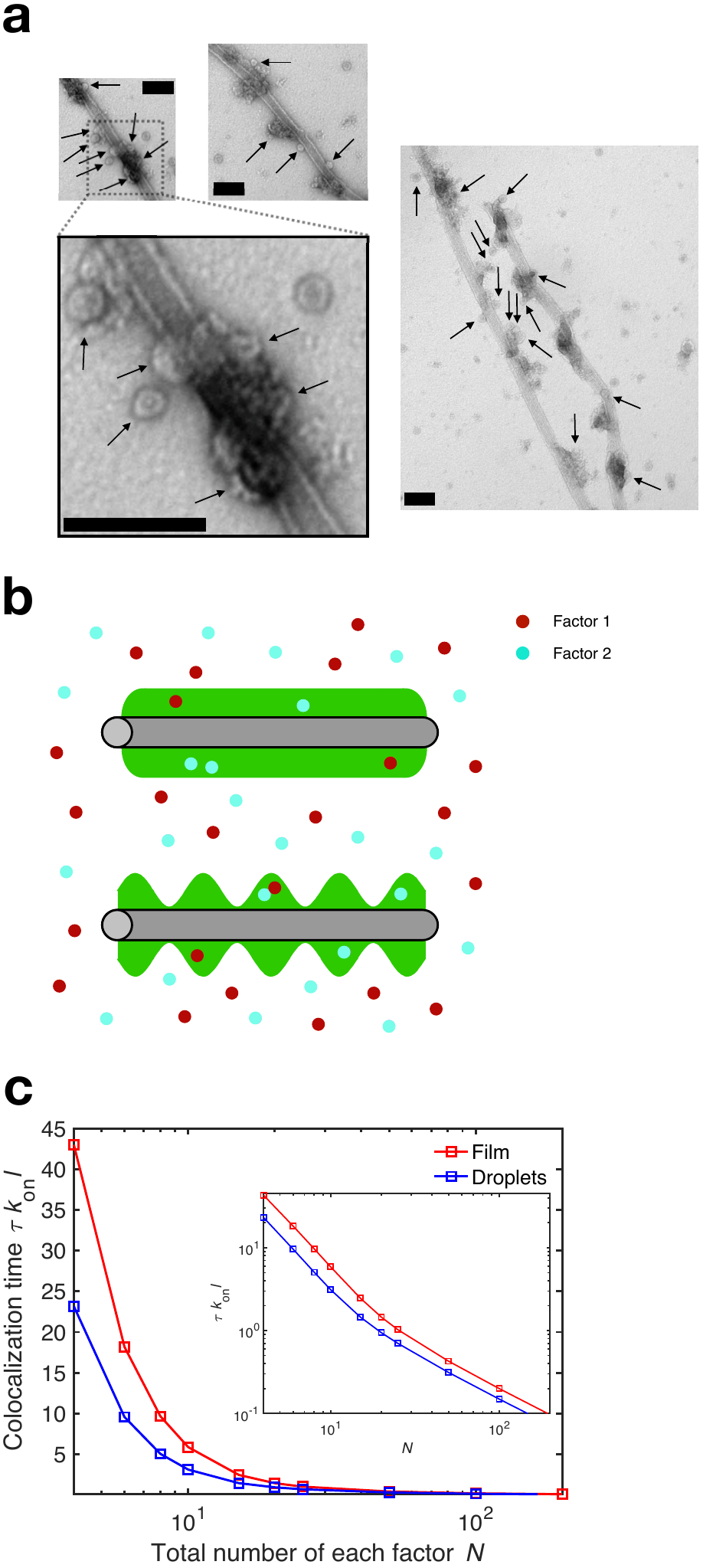}} 
\end{figure}

How might TPX2 droplets facilitate branching microtubule nucleation? Noting that this process requires the coordinated action in time and space of at least two additional factors, augmin and $\gamma$-TuRC, we first imaged the localization of $\gamma$-TuRC on microtubules in the presence of TPX2 and augmin using electron microscopy. We found that the ratio of $\gamma$-TuRC on microtubules to $\gamma$-TuRC on the grid surface was $0.05 \pm 0.05$ without TPX2 and augmin (mean $\pm$ standard deviation, $N=3$ microtubules, \hyperref[fig:s8]{Supplementary Fig.~8}, \hyperref[table:s4]{Supplementary Table 4}). With TPX2 and augmin, this ratio was $0.48 \pm 0.04$ (mean $\pm$ standard deviation, $N=4$ microtubules, \hyperref[table:s5]{Supplementary Table 5}), confirming that TPX2 and augmin preferentially localize $\gamma$-TuRC to microtubules. We observed that multiple $\gamma$-TuRCs cluster inside TPX2 droplets spaced $0.25\pm0.09$ $\mu$m (mean $\pm$ standard deviation, $N=4$ microtubules) apart along microtubules (\hyperref[fig:4]{Fig.~4a}), consistent with a recent report \cite{Alfaro-Aco2019}. The ratio between the number of $\gamma$-TuRCs inside TPX2 droplets to the number on bare regions of the same microtubules was $4.8\pm2.0$ (mean $\pm$ standard deviation, $N=3$ microtubules). Although this is an underestimate, given the difficulty of counting $\gamma$-TuRCs in TPX2 droplets, these results demonstrate that $\gamma$-TuRC preferentially localizes to TPX2 droplets along microtubules.

The first step in branching is the binding of TPX2 to the microtubule, which then localizes the other factors \cite{Thawani2019}. As such, we hypothesized that regularly spaced TPX2 droplets lead to more efficient colocalization of factors than a uniform coating (\hyperref[fig:4]{Fig.~4b}). For a uniform coating, multiple factors must search a greater length along the microtubule before finding each other to nucleate a new branch. With regularly spaced droplets, the explored distance is shorter, which reduces the search time. We performed kinetic Monte Carlo simulations \cite{Gillespie1977} for two factors binding to (with rate $k_{\mathrm{on}}$) and unbinding from (with rate $k_{\mathrm{off}}$) a microtubule of length $l$ with a uniform TPX2 coating and a periodic pattern of TPX2 droplets (\hyperref[theory:1-3]{Supplementary Theory}). These results show that the time $\tau$ to colocalize on the microtubule, and hence the minimum time for nucleation of a new branch, is smaller for droplets than for a uniform layer (\hyperref[fig:4]{Fig.~4c}, \hyperref[fig:s9]{Supplementary Fig.~9}). As a negative control for this model, we used AFM to measure the topography of a C-terminal fragment of TPX2 on microtubules. This fragment is known to be less efficient at nucleating branches in cytosol \cite{King2019}. Consistent with our model, it did not form droplets on microtubules (\hyperref[fig:s7]{Supplementary Fig.~7a}). Thus, synergistic with TPX2's ability to recruit tubulin \cite{King2019} and its high concentration as a condensate, its organization into droplets partitions the microtubule so that multiple factors can more easily find each other. Taken together, TPX2's phase behaviour enhances reaction kinetics via droplet patterning, condensate concentration, and tubulin recruitment.

It is important to think about our model in a cellular context. During cell division, TPX2 is released as a gradient in the vicinity of chromosomes \cite{caudron2005spatial}. The typical TPX2 concentration in \textit{X. laevis} is $90$ nM \cite{wuhr2014deep} and the typical gradient length is $\sim10$ $\mu$m \cite{caudron2005spatial}. This gives $\sim2\times10^5$ TPX2 molecules that are available to condense on microtubules near chromosomes, assuming a spherical volume. We estimate the concentration of TPX2 in the condensed phase to be $10^4~\mu\mathrm{m}^{-3}$ using our AFM data (\hyperref[methods:concentrationEstimation]{Methods}). There are then $\sim80$ TPX2 molecules needed to form a 10-nm-high condensed film on a typical $7$-$\mu$m-long, 25-nm-diameter microtubule \cite{brugues2012nucleation}. Therefore, TPX2 can coat $\sim3\times10^3$ microtubules during cell division. Given that the density of microtubules in the metaphase spindle is $\sim2$ $\mu \mathrm{m}^{-3}$ within $10$ $\mu \mathrm{m}$ of chromosomes \cite{brugues2012nucleation}, $\sim8\times10^3$ microtubules lie in the vicinity of chromosomes. Thus, TPX2 can coat $\sim40 \%$ of the metaphase microtubule mass near chromosomes at this film thickness. We hypothesize that the Rayleigh-Plateau instability is most relevant during early spindle assembly in order to accelerate the generation of microtubules, as TPX2  is responsible for creating most of the spindle microtubules via branching nucleation \cite{decker2018autocatalytic}, in particular during early stages of spindle assembly \cite{petry2011augmin}.

As the study of liquid-like protein condensates has intensified \cite{Shin2017, alberti2019considerations}, the physical phenomenology has been dominated by optical observations of droplets in solution or on microtubules. Here, we quantitatively demonstrate the emergence of non-trivial hydrodynamic features such as films and spatiotemporal periodic instabilities that arise when a condensate interacts with a filament. In future work, it will be interesting to explore how multiple proteins, such as TPX2 and BuGZ \cite{Jiang2015}, condense on the microtubule, as multi-protein condensates in solution have been reported \cite{feric2016coexisting}. We suspect that interfacial physics could manifest itself in other ways when condensates interact with cellular filaments, such as via elastocapillary effects  \cite{Duprat2012} that could produce forces between cytoskeletal filaments or other semi-flexible macromolecules such as RNA and DNA \cite{li2019llpsdb}.

\newpage

\section*{Acknowledgements}

We thank Drs.~Stephanie Lee, Tseng-Ming Chou, and Matthew Libera at Stevens Institute of Technology for access to their AFM; Drs.~Ian Armstrong and Samrat Dutta at Bruker for access to and support for their AFM; Drs.~Matthew King, Benjamin Bratton, Mohammad Safari, Matthias Koch, Pierre Ronceray, and Ned Wingreen for helpful discussions; Dr.~Akanksha Thawani for purification of TPX2; Henry Ando, Caroline Holmes, physiology students Dr.~Valentina Baena, Davis Laundon, and Linda Ma, and the Physiology Course at the Marine Biological Lab for assisting with the first AFM trials; and Princeton’s Imaging and Analysis Center, which is partially supported by the Princeton Center for Complex Materials, an NSF-MRSEC program (DMR-1420541).

B.G.~was supported by PD Soros and NSF GRFP. S.U.S.~was supported by NIH NCI NRSA 1F31CA236160 and NHGRI training grant 5T32HG003284. This work was funded by NIH NIA 1DP2GM123493, Pew Scholars Program 00027340, Packard Foundation 2014–40376, and CPBF NSF PHY-1734030. 

\section*{Author contributions}

S.U.S., B.G., J.W.S., H.A.S., and S.P. conceptualized the project. B.G.~and S.U.S.~performed fluorescence microscopy, and B.G.~performed analysis of fluorescence microscopy data. B.G.~performed TPX2-only electron microscopy with assistance from R.A.~and S.U.S., theory, and simulations. S.U.S.~performed atomic force microscopy with assistance from B.G., associated data analysis, and meiotic cytosol experiments. R.A.~performed branching reconstitution and multiple-protein electron microscopy, and S.U.S.~performed analysis of branching reconstitution data. S.U.S.~and B.G.~wrote the paper with assistance from J.W.S., H.A.S, and S.P. J.W.S., H.A.S., and S.P. supervised the research. All authors discussed and interpreted results and revised the paper.

\section*{Competing interests}

The authors declare no competing interests.

\section*{Ethics}

Data, code, and materials used are available upon request.

Animal care was done in accordance with recommendations in the Guide for the Care and Use of Laboratory Animals of the NIH and the approved Institutional Animal Care and Use Committee (IACUC) protocol 1941-16 of Princeton University.

\newpage

\vspace*{8mm}

\begin{center}
\LARGE{A hydrodynamic instability drives protein droplet formation on microtubules to nucleate branches: supplementary information} \\
\end{center}

\begin{center}
\large{Sagar U. Setru$^{\dag, 1}$, Bernardo Gouveia$^{\dag, 2}$, Raymundo Alfaro-Aco$^3$, Joshua W. Shaevitz$^{\ast,1,4}$, Howard A. Stone$^{\ast, 5}$, Sabine Petry$^{\ast, 3}$}
\end{center}

\begin{center}
$^{\dag}$These authors contributed equally. \\ {\parbox{\linewidth}{ \normalsize{$^{\ast}$To whom correspondence should be addressed; E-mail: \mbox{spetry@princeton.edu}, \mbox{hastone@princeton.edu}, \mbox{shaevitz@princeton.edu}.}}}
\end{center}

\newpage

\section*{Supplementary Figures}

\begin{figure}[h!]
\centering
\includegraphics[width=16cm]{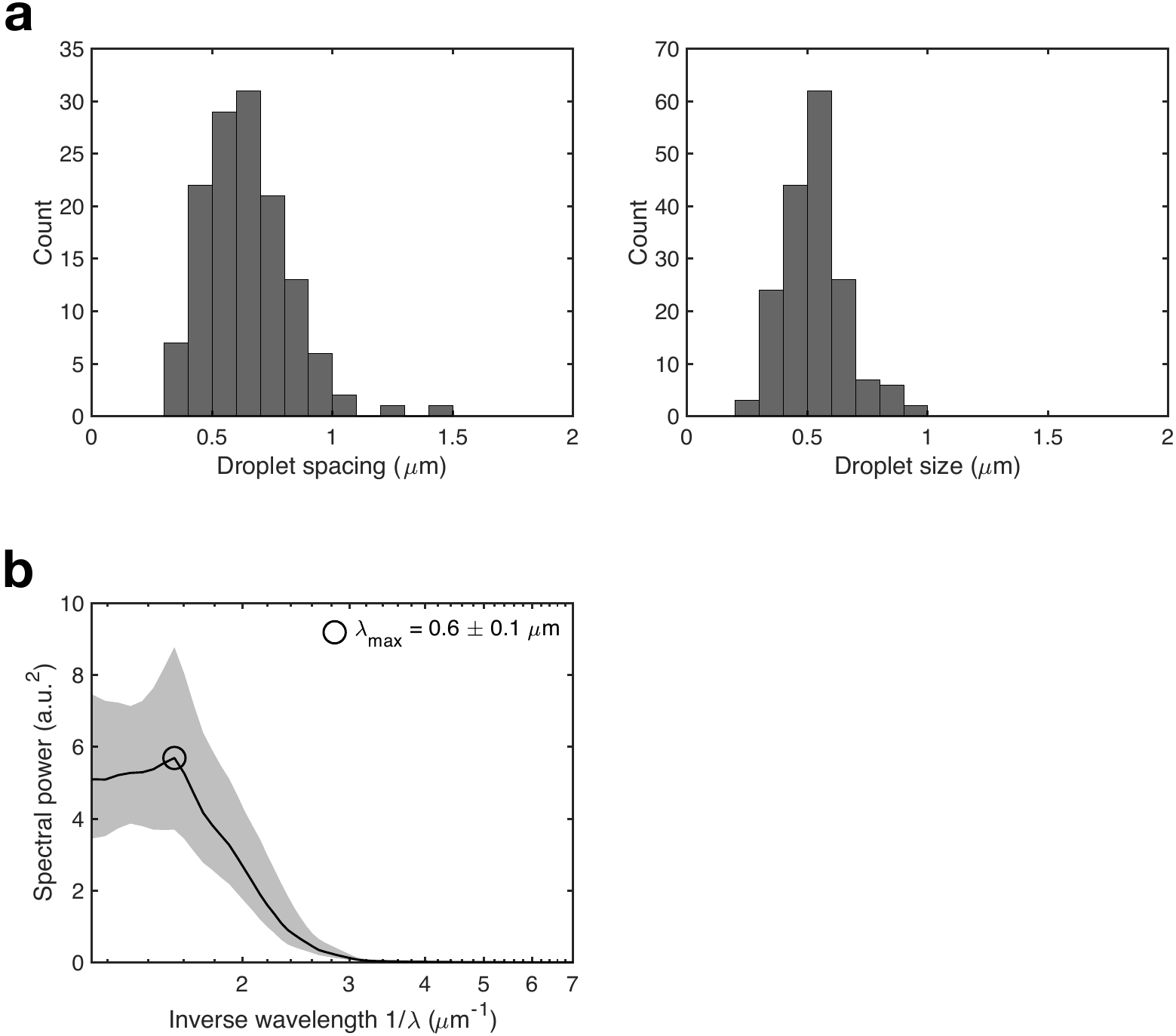}
\caption*{\textbf{Supplementary Fig. 1: Statistics of droplet patterned microtubules imaged with TIRF microscopy.} (a) Histogram of droplet sizes and spacings for TIRF experiments at 1 $\mu$M GFP-TPX2. $N=35$ microtubules were analyzed with a mean size of $0.5 \pm 0.1~\mu\mathrm{m}$ and spacing of $0.6 \pm 0.2~\mu\mathrm{m}$ (mean $\pm$ standard deviation). (b) Average power spectrum of GFP-TPX2 fluorescence intensities of droplet patterns for TIRF experiments at 1 $\mu$M GFP-TPX2 ($N=35$ microtubules). The peak indicates the emergence of a periodic pattern with wavelength $\lambda_{\mathrm{max}} = 0.6 \pm 0.1~\mu\mathrm{m}$ (mean $\pm$ standard deviation), in agreement with the histogram analysis. The shaded region represents 95\% bootstrap confidence intervals.} \label{fig:s1}

\end{figure}

\clearpage

\begin{figure}[h]
\centering
\floatbox[{\capbeside\thisfloatsetup{capbesideposition={right,top},capbesidewidth=5.8cm}}]{figure}[\FBwidth]
{\caption*{\textbf{Supplementary Fig. 2: TPX2 on the microtubule can appear uniform when imaged via optical microscopy.} (a) TIRF microscopy time lapses showing that a 0.1~$\mu$M TPX2 coating does not break up into visible droplets like the 1~$\mu$M TPX2 coating does. (b) Branching microtuble nucleation visualized by TIRF microscopy in \textit{X. laevis} meiotic cytosol at 0.1~$\mu$M TPX2, indicating that branching can occur from diffraction limited droplets.}\label{fig:s2}}
{\includegraphics[width=10cm]{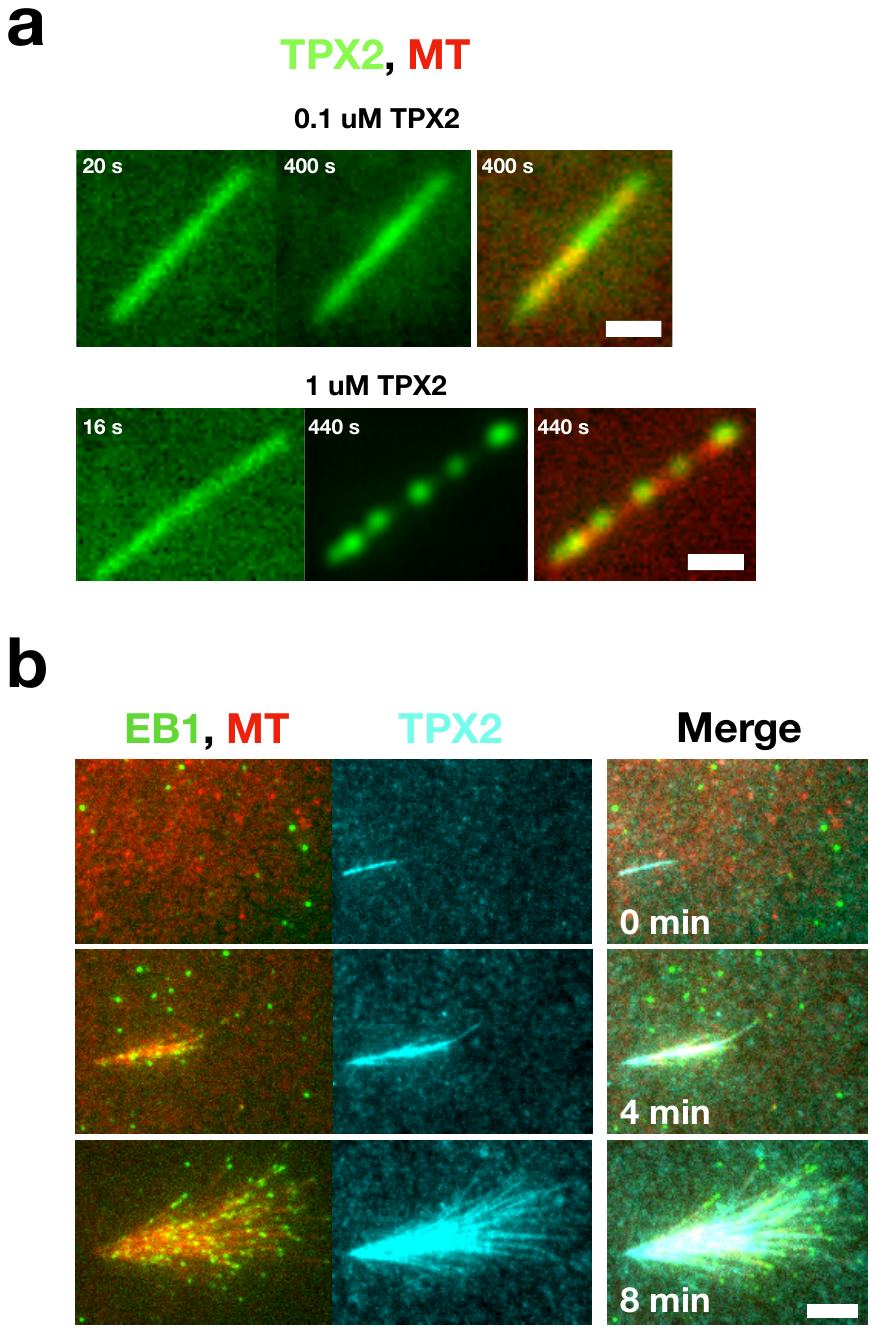}}
\end{figure}
\clearpage
\begin{figure}[h!]
\centering
\floatbox[{\capbeside\thisfloatsetup{capbesideposition={right,top},capbesidewidth=6cm}}]{figure}[\FBwidth]
{\caption*{\textbf{Supplementary Fig. 3: Raw and smoothed AFM height profiles, and power spectra of raw height profiles.}
  (a) Raw height profiles of the topographies in \hyperref[fig:2]{Fig.~2a}. The smoothed profile from \hyperref[fig:2]{Fig.~2b} is shown again for reference. (b) Power spectra of the raw height profiles in (a). The red curve shows the mean $\pm$ standard error of the mean over nine topographies of the microtubule after the droplet pattern had formed. The frequency $f$ at which the peak in the red curve occurs gives the droplet spacing measured for this microtubule, according to $\lambda=1/f$.
  }
  }
  {\includegraphics[width=10cm]{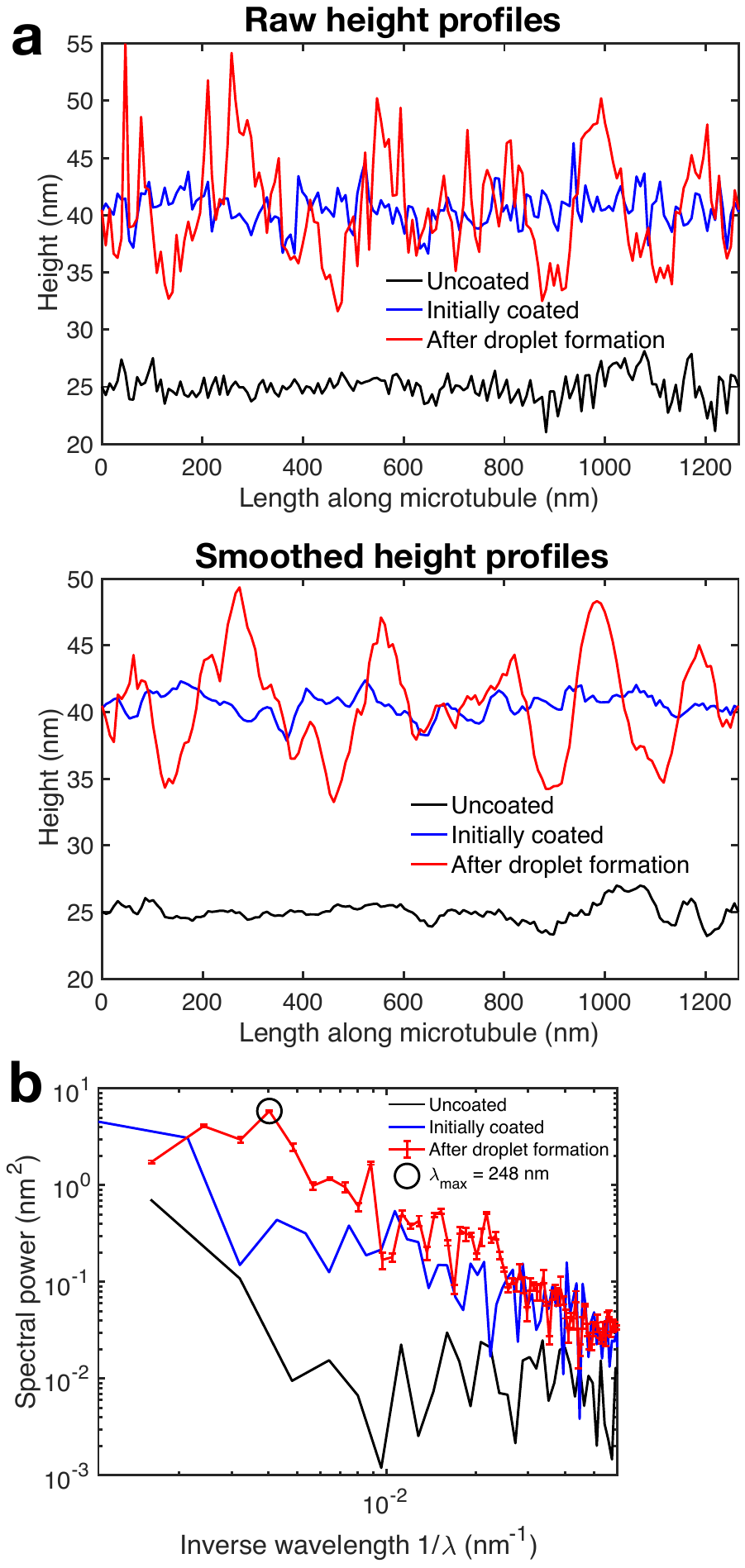}}
  \label{fig:s3}
  \end{figure}
\clearpage
\begin{figure}
    \includegraphics[]{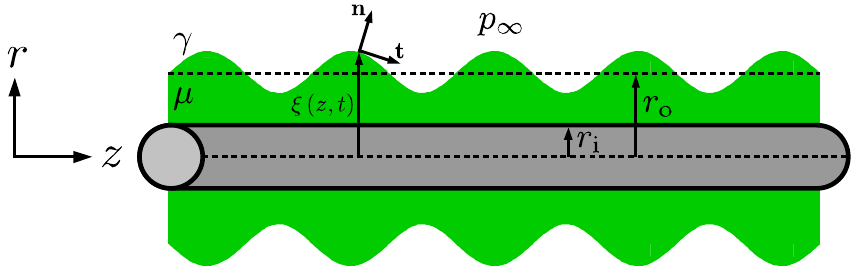}
  \caption*{\textbf{Supplementary Fig. 4: The Rayleigh-Plateau instability.} The viscosity of the condensed film is $\mu$, $\gamma$ is the surface tension of the interface, and $p_{\infty}$ is the far field pressure provided by the solvent. The microtubule has radius $r_{\mathrm{i}}$. Initially, the interface is flat at $\xi\left(z,t=0\right) = r_{\mathrm{o}}$, but this scenario is unstable against the capillary pressure $\gamma/r_{\mathrm{o}}$, so $\xi\left(z,t\right)$ will evolve to a lower energy state. The unit normal $\textbf{n}$ and unit tangent $\textbf{t}$ track the geometry of the interface during its evolution.}
 \label{fig:s4}
\end{figure}
\clearpage
\begin{figure}[h]
    \includegraphics[width=5.5in, height=6.2in]{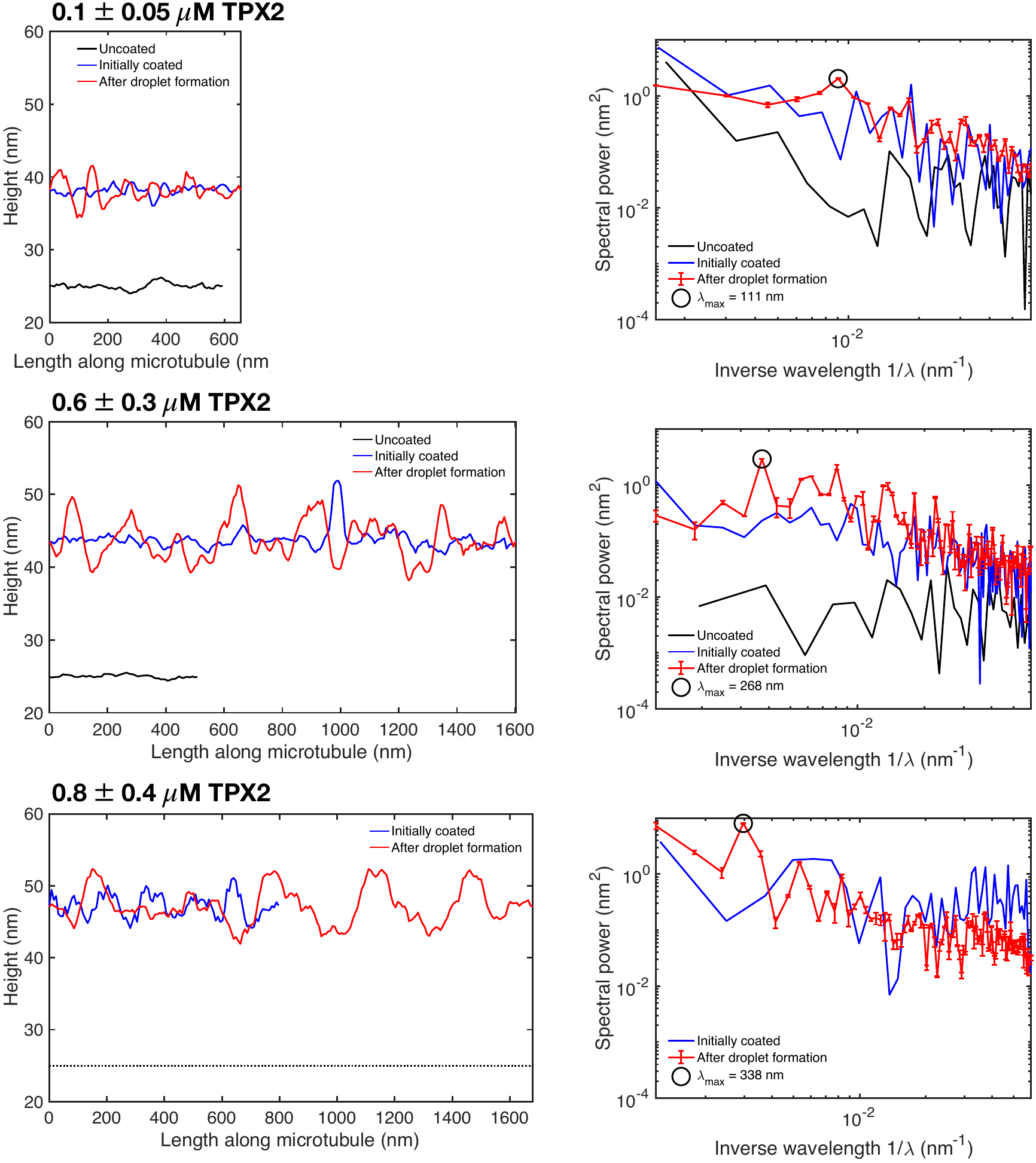}
  \caption*{\textbf{Supplementary Fig. 5: AFM height profiles and power spectra at additional TPX2 concentrations.} For $0.1\pm0.05~\mu\mathrm{M}$, the power spectrum is averaged over $N=5$ topographies after the droplet pattern had formed. For $0.6\pm0.3~\mu\mathrm{M}$, $N = 3$. For $0.8\pm0.4~\mu\mathrm{M}$, $N = 4$; the uncoated height profile for this specific microtubule is unavailable because the sample moved after TPX2 addition. Height profiles were smoothed using a moving-average window of $40~\mathrm{nm}$; power spectra were taken from raw height profiles. All power spectra after droplet formation show  mean $\pm$ standard  error  of  the  mean.} 
 \label{fig:s5}
\end{figure}
\clearpage
\begin{figure}[h!]
\centering
\floatbox[{\capbeside\thisfloatsetup{capbesideposition={right,top},capbesidewidth=6cm}}]{figure}[\FBwidth]
{\caption*{\textbf{Supplementary Fig. 6: Growth of the condensed film.} (a) Schematic of the model for growth of the condensed protein film. Microtubules of radius $r_{\mathrm{i}}$ are spaced periodically by a distance $2\bar{R}$, where $\bar{R} = 1/\sqrt{\pi n l}$ where $l$ is the typical microtubule length and $n$ is the number density of microtubules. Soluble protein with diffusivity $D$ phase separates from solution and nucleates a spatially uniform condensed film on the microtubule surface, whose interfacial position we denote by $r = \xi\left(t\right)$. (b) Final film thickness $h$ versus initial concentration $c_{0}$ as measured by atomic force microscopy (blue) and as predicted by equation (\ref{eq:final_film}) (black), using $1/c_{\mathrm{R}}\left(\bar{R}^2/r_{\mathrm{i}}^2 - 1\right)$ as a least-squares fit parameter. (c) Evolution of the interfacial position of the film $\xi/r_{\mathrm{i}}$ over time $\pi n l D t$ for $S=r_{\mathrm{i}}/r_{\mathrm{o}}\in\left[0.5, 0.7\right]$, which is our experimentally observed range of $S$. Solid lines are the exact solution and dashed lines are the asymptotic formula (\hyperref[eq:late-time-solution]{34b}).}
\label{fig:s6}}
{\includegraphics[width=8cm]{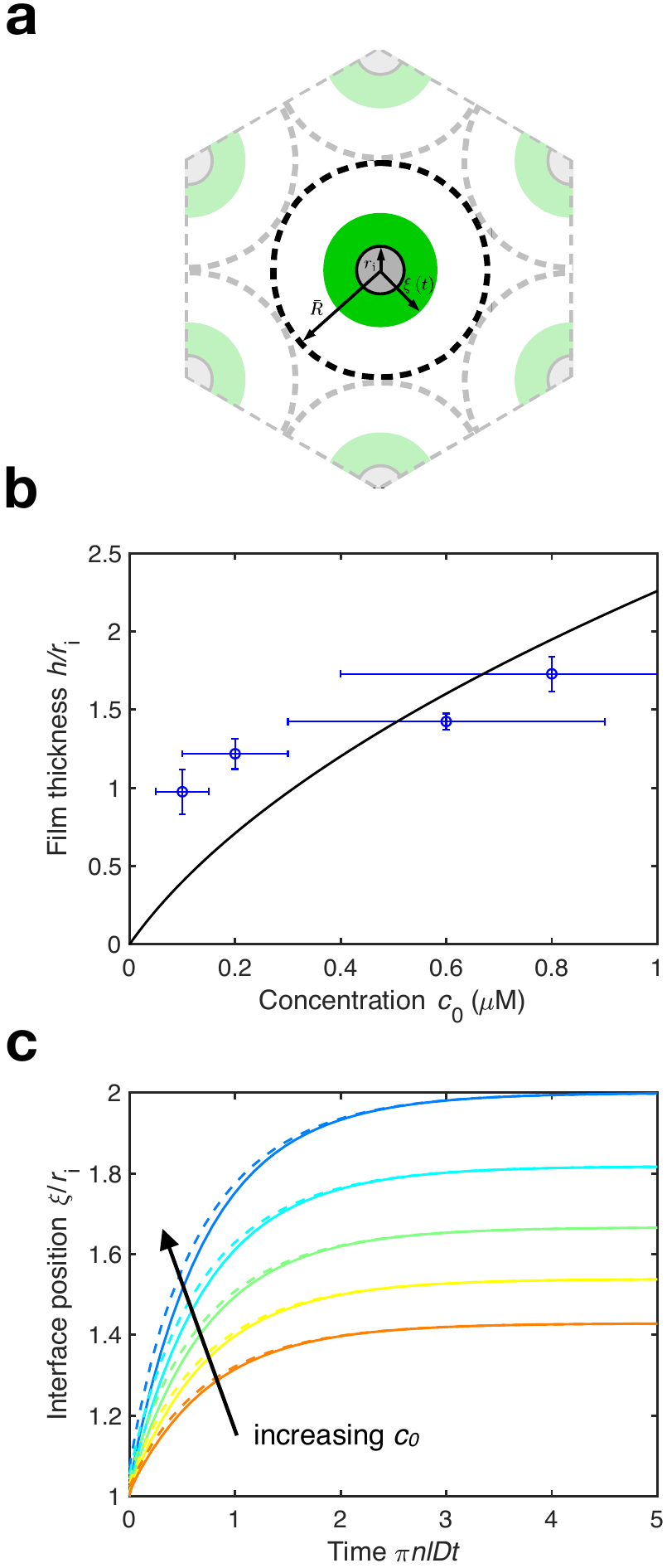}}
\end{figure}
\clearpage
\begin{figure}[h]
    \includegraphics[width=4.5in, height=4.5in]{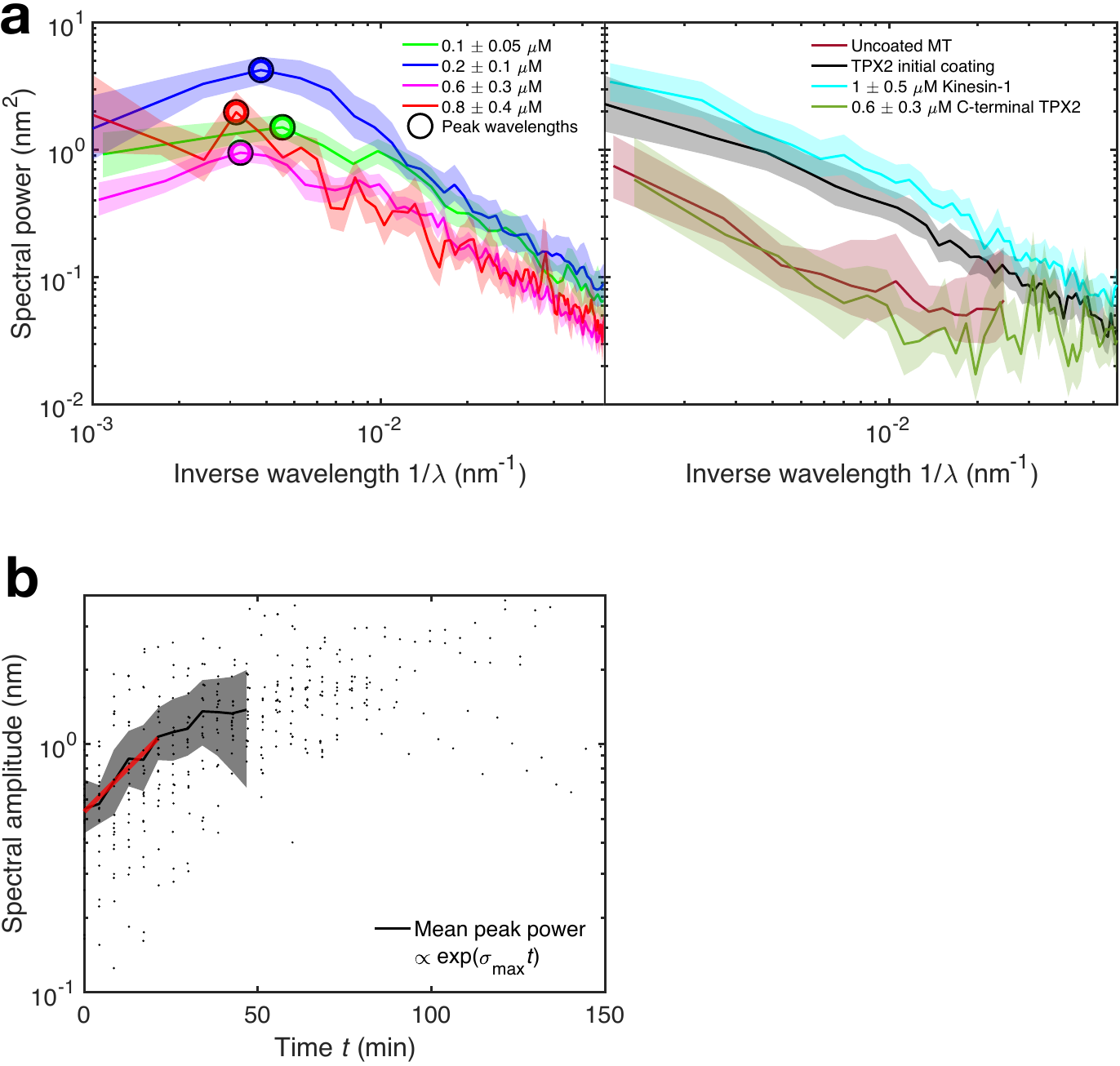}
  \caption*{\textbf{Supplementary Fig. 7: Average power spectra from AFM data for all concentrations of TPX2 and for uncoated, initially TPX2-coated, C-terminal-TPX2-bound, and kinesin-1-bound microtubules; the growth of the instability for early times is exponential.} (a) Peaks indicate characteristic wavelengths that correspond to a typical droplet spacing (\hyperref[table:s3]{Supplementary Table 3}, $N=$~18, 17, 14, and 8 microtubules, respectively, for increasing TPX2 concentration). Also included are average power spectra for uncoated microtubules ($N = 29$ microtubules), microtubules initially coated uniformly with TPX2 ($N = 25$ microtubules), kinesin-bound microtubules ($N = 19$ microtubules), and C-terminal-TPX2-bound microtubules ($N = 4$ microtubules)---none of which show any characteristic spatial features. For kinesin-bound microtubules, $h=2.9 \pm 2.0~\mathrm{nm}$, consistent with what one would expect for the kinesin construct used \cite{cao2014structure, shang2014high}. For C-terminal-TPX2-bound microtubules, $h=3.7 \pm 1.8~\mathrm{nm}$. Heights are mean $\pm$ standard deviation. Shaded regions represent 95\% bootstrap confidence intervals. (b) The average spectral amplitude (black line, $N=21$ microtubules) at the most unstable frequency grows exponentially for early times. Spectral amplitude = $\sqrt{\mathrm{spectral~power}}$. Individual measurements are black dots. The shaded region represents $95\%$ bootstrap confidence intervals. At later times, the spectral amplitude levels off due to nonlinear forces as the pattern sets in. For the exponential fit (red line), $\sigma_{\mathrm{max}} = 0.03~\mathrm{min}^{-1}$, with $R^{2} = 0.75$.}
 \label{fig:s7}
\end{figure}
\clearpage
\begin{figure}[h!]
\centering
\floatbox[{\capbeside\thisfloatsetup{capbesideposition={right,top},capbesidewidth=5.8cm}}]{figure}[\FBwidth]
    {\caption*{\textbf{Supplementary Fig. 8: $\gamma$-TuRC localization on bare microtubules.} Typical electron microscopy experiment with just $\gamma$-TuRC and microtubules. No TPX2 or augmin is present. The localization of $\gamma$-TuRC to the microtubule without TPX2 and augmin is negligible (\hyperref[table:s4]{Supplementary Table 4}). Scale bar is 100 nm.}\label{fig:s8}}
{\includegraphics[width=8cm]{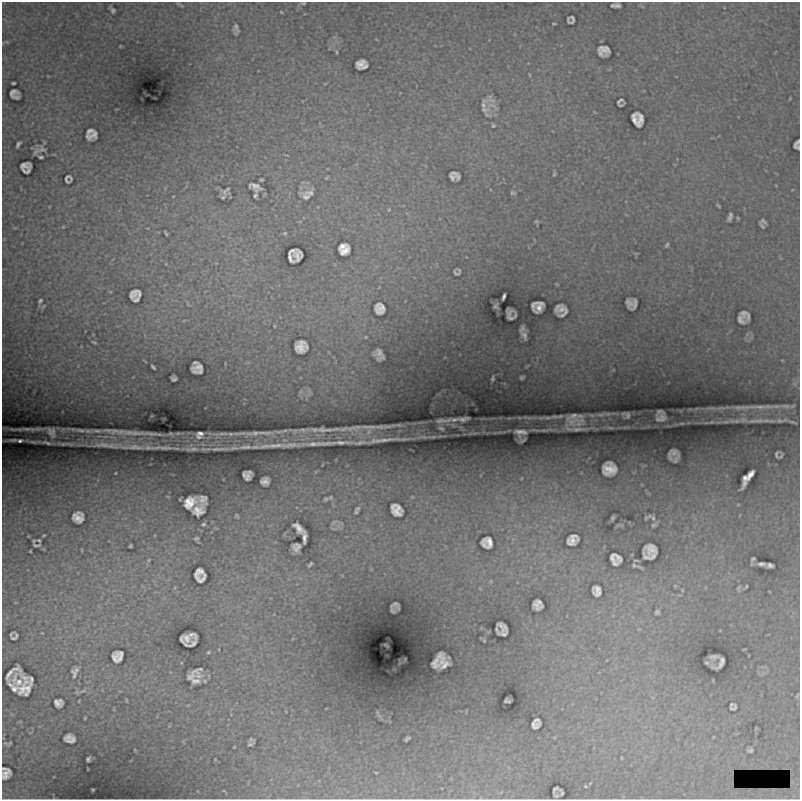}}
\end{figure}
\clearpage
\begin{figure}[h!]
\centering
\floatbox[{\capbeside\thisfloatsetup{capbesideposition={right,top},capbesidewidth=5.8cm}}]{figure}[\FBwidth]
{\caption*{\textbf{Supplementary Fig. 9: Parametric study of Monte Carlo simulations.} (a) Time $\tau$ to colocalize two distinct factors, and hence the minimum time to form a branch, as a function of the total number of each factor $N$ and $s=k_{\mathrm{on}}l/k_{\mathrm{off}}$ for a uniform and periodic protein coating. For a given $s$, the periodic coating is always more efficient at colocalizing well-mixed factors. Each data point is the average of $10^{7}$ independent simulations. (b) Typical histogram of $10^{7}$ independent simulations for two factors with $N=50$ and $s=10$.}\label{fig:s9}}
{\includegraphics[width=8cm]{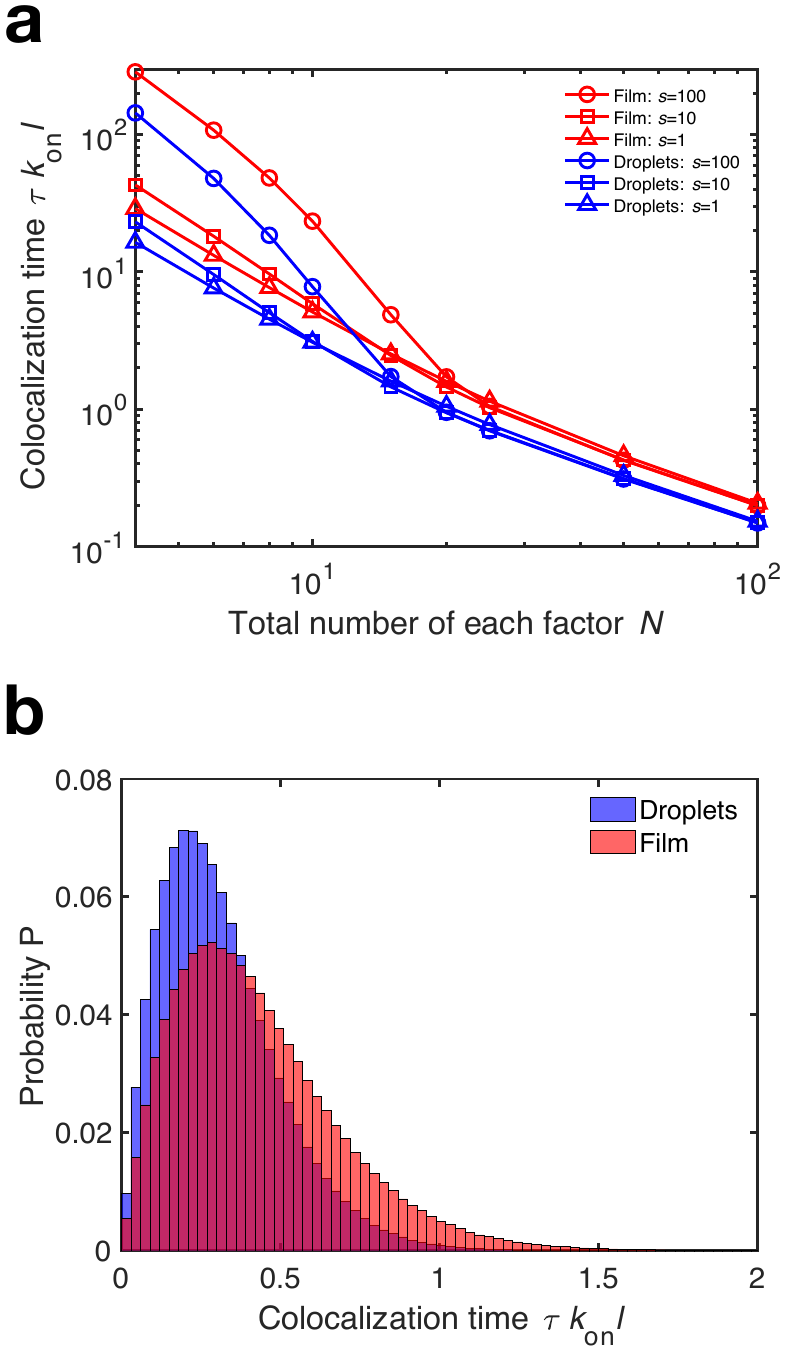}}
\end{figure}

\newpage
\pagebreak

\section{Theory and simulation}

\subsection{Rayleigh-Plateau instability} \label{theory:1-1}

Assuming conditions are met for the nucleation and growth of the condensed film to be time scale separated from the subsequent Rayleigh-Plateau instability (\hyperref[theory:1-2]{Supplementary Information 1.2}), we proceed to analyze the stability of a condensed film of thickness $h$ on a microtubule of radius $r_{\mathrm{i}}$ (\hyperref[fig:s4]{Supplementary Fig.~4}). The nanometric length scales of the system are such that inertial effects are completely negligible, and so our starting point is the Stokes equations
\begin{linenomath*}
\begin{subequations} \label{eq:stokes}
	\begin{eqnarray}
	\boldsymbol{\nabla}\cdot\boldsymbol{u} &=& 0 \\
	\mu\nabla^2\boldsymbol{u} &=& \boldsymbol{\nabla}p
	\end{eqnarray}
\end{subequations}
\end{linenomath*}
for the velocity $\boldsymbol{u}$ and pressure $p$ fields in the condensate, where $\mu$ is the viscosity of the condensate. This simplified starting point is the only difference between our derivation and that in \cite{Goren1962}.

For the boundary conditions at $r=r_{\mathrm{i}}$ we require no-slip on the microtubule surface
\begin{linenomath*}
\begin{equation} \label{eq:no_slip}
\boldsymbol{u}\left(r=r_{\mathrm{i}}, z\right) = \boldsymbol{0}.
\end{equation}
\end{linenomath*}
At the interface $r=\xi\left(z,t\right)$ between the film and the surrounding fluid, we assume that the viscosity of the solvent $\mu_{\mathrm{s}}$ is much less than $\mu$, which is an excellent assumption for condensed proteins \cite{Taylor2016}. Hence, we can write for the tangential stress condition
\begin{linenomath*}
\begin{equation} \label{eq:tangential}
\textbf{t}\cdot\boldsymbol{\tau}\cdot\textbf{n} = 0\mathrm{\ at\ }r=\xi\left(z,t\right),
\end{equation}
\end{linenomath*}
where $\boldsymbol{\tau}=\mu\left(\boldsymbol{\nabla}\boldsymbol{u} + \boldsymbol{\nabla}\boldsymbol{u}^{\mathrm{T}}\right)$ is the viscous stress tensor. $\textbf{n}=\left(\textbf{e}_{r} - \xi'\textbf{e}_{z}\right)/\sqrt{1 + {\xi'}^{2}}$ is the unit normal and $\textbf{t}=\left(\xi'\textbf{e}_{r} + \textbf{e}_{z}\right)/\sqrt{1 + {\xi'}^{2}}$ is the unit tangent expressed in local cylindrical coordinates with $\textbf{e}_{\alpha}\cdot\textbf{e}_{\beta}=\delta_{\alpha\beta}$, where $\xi'=\partial_{z}\xi$. When the interface is crossed, there is a Young-Laplace jump in the normal stress due to the surface tension $\gamma$ of the form
\begin{linenomath*}
\begin{equation} \label{eq:normal}
-p_{\infty} - \textbf{n}\cdot\boldsymbol{\sigma}\cdot\textbf{n} = \gamma\boldsymbol{\nabla}\cdot\textbf{n}\mathrm{\ at\ }r=\xi\left(z,t\right),
\end{equation}
\end{linenomath*}
where $\boldsymbol{\sigma}=-p\textbf{I}+\boldsymbol{\tau}$ is the total stress. Lastly, we account for the kinematic condition that the interface advects with the flow
\begin{linenomath*}
\begin{equation} \label{eq:kinematic}
\left(\partial_{t} + \boldsymbol{u}\cdot\boldsymbol{\nabla}\right)\left(r - \xi\right)=0\mathrm{\ at\ }r=\xi\left(z,t\right),
\end{equation}
\end{linenomath*}
which closes the problem.

To make progress, we assume that all motions are axisymmetric by the symmetry of the cylindrical geometry, so that $u_{\phi}=0$ and $\partial_{\phi}\mapsto0$. In this case we can introduce the stream function $\psi$ by $u_{r}=\left(1/r\right)\partial_{z}\psi$ and $u_{z} = -\left(1/r\right)\partial_{r}\psi$, which automatically satisfies equation (\hyperref[eq:stokes]{1a}). Equations (\hyperref[eq:stokes]{1b}) simplify to
\begin{linenomath*}
\begin{subequations} \label{eq:axisymmetric-stokes}
\begin{eqnarray}
\partial_{r}p &=& \mu\left(\partial_{r}^2 + \frac{1}{r}\partial_{r} - \frac{1}{r^2} + \partial_{z}^2\right)\frac{1}{r}\partial_{z}\psi \\
\partial_{z}p &=& -\mu\left(\partial_{r}^2 + \frac{1}{r}\partial_{r} + \partial_{z}^2\right)\frac{1}{r}\partial_{r}\psi.
\end{eqnarray}
\end{subequations}
\end{linenomath*}
Taking $\partial_{z}$ of equation (\hyperref[eq:axisymmetric-stokes]{6a}) and equating it to $\partial_{r}$ of equation (\hyperref[eq:axisymmetric-stokes]{6b}) furnishes, after simplification, the fourth-order equation 
\begin{linenomath*}
\begin{equation} \label{eq:field-equation}
D^2\psi=0,
\end{equation}
\end{linenomath*}
where $D=\partial_{r}^2-\left(1/r\right)\partial_{r}+\partial_{z}^2$.

The equilibrium solutions that satisfy equation (\ref{eq:field-equation}) and boundary conditions (\ref{eq:no_slip}-\ref{eq:kinematic}) are $\psi=0$, $p=p_{\infty}+\gamma/r_{\mathrm{o}}$, and $\xi=r_{\mathrm{o}}$. We now perform a linear stability analysis about these equilibrium fields by positing the expansions
\begin{linenomath*}
\begin{equation} \label{eq:expansions}
\left(\psi, p, \xi\right) = \left(0, p_{\infty}+\gamma/r_{\mathrm{o}}, r_{\mathrm{o}}\right) + \sum_{k}\left(\hat{\psi}_{k}\left(r\right), \hat{p}_{k}\left(r\right),\hat{\xi}_{k}\right)e^{\sigma t + ikz},
\end{equation}
\end{linenomath*}
where we interpret the hatted quantities as \textit{small} perturbations to the equilibrium solutions, and hence we will retain at most up to linear order in hatted quantities. Therefore, by linearity, we can conduct the analysis for each mode $k$ separately, with the end goal being to arrive at the dispersion relation $\sigma = \sigma\left(k\right)$.

Substituting equation (\ref{eq:expansions}) into equation (\ref{eq:field-equation}) results in
\begin{linenomath*}
\begin{equation} \label{eq:fourier_field_equation}
D_{k}^2\hat{\psi}_{k} = 0,
\end{equation}
\end{linenomath*}
where $D_{k} = \mathrm{d}^2/\mathrm{d}r^2 - \left(1/r\right)\mathrm{d}/\mathrm{d}r - k^2$. To solve equation (\ref{eq:fourier_field_equation}), we construct the solution $\hat{\psi}_{k}=\hat{\psi}_{k}^{(1)} + \hat{\psi}_{k}^{(2)}$, where $D_{k}\hat{\psi}_{k}^{(1)}=0$ and $D_{k}\hat{\psi}_{k}^{(2)}=\hat{\psi}_{k}^{(1)}$. Making the substitution $\hat{\psi}_{k}^{(1)}=r\zeta\left(r\right)$ transforms $D_{k}\hat{\psi}_{k}^{(1)}=0$ into a modified Bessel equation of order $1$ for $\zeta\left(r\right)$, so the solution is $\hat{\psi}_{k}^{(1)}=ArI_{1}\left(kr\right)+BrK_{1}\left(kr\right)$, where $I_{p}$ and $K_{p}$ are modified Bessel functions of the first and second kind of order $p$, respectively. We can now write $D_{k}\hat{\psi}_{k}^{(2)}=rI_{1}\left(qr\right) + rK_{1}\left(qr\right)$, where we will let $q\rightarrow k$. The particular solution is found by inspection to be $\hat{\psi}_{k}^{(2)}=\left[rI_{1}\left(qr\right) + rK_{1}\left(qr\right)\right]/\left(q^2-k^2\right)$. We perform the necessary limit by letting $q = k +\epsilon$ and taking $\epsilon\rightarrow0$. Extracting the only surviving linearly independent terms, while regularizing integration constants where necessary, furnishes the particular solution $\hat{\psi}_{k}^{(2)} = r^{2}I_{2}\left(kr\right) + r^{2}K_{2}\left(kr\right)$. Hence
\begin{linenomath*}
\begin{equation} \label{eq:fourier_stream_function}
\hat{\psi}_{k}=ArI_{1}\left(kr\right) + BrK_{1}\left(kr\right) + Cr^{2}I_{2}\left(kr\right) + Dr^{2}K_{2}\left(kr\right),
\end{equation}
\end{linenomath*}
where the integration constants $A,B,C$ and $D$ are to be determined by the boundary conditions (\ref{eq:no_slip}-\ref{eq:kinematic}).

It is now convenient to define the dimensionless variables $K=kr_{\mathrm{o}}$, $\Sigma=\mu r_{\mathrm{o}}\sigma/\gamma$, and $S=r_{\mathrm{i}}/r_{\mathrm{o}}$. We substitute the stream function (\ref{eq:fourier_stream_function}) into the no-slip condition (\ref{eq:no_slip}), resulting in
\begin{linenomath*}
\begin{subequations} \label{eq:fourier_no_slip}
\begin{eqnarray}
AI_{1}\left(KS\right) + BK_{1}\left(KS\right) + 
\tilde{C}I_{2}\left(KS\right) + \tilde{D}K_{2}\left(KS\right) &=& 0 \\
AI_{0}\left(KS\right) - BK_{0}\left(KS\right) + 
\tilde{C}I_{1}\left(KS\right) - \tilde{D}K_{1}\left(KS\right) &=& 0,
\end{eqnarray}
\end{subequations}
\end{linenomath*}
where $\tilde{C}=Cr_{\mathrm{o}}$ and $\tilde{D}=Dr_{\mathrm{o}}$. Substituting the expansion (\ref{eq:expansions}) into the tangential stress condition (\ref{eq:tangential}) while retaining only the leading-order terms, then using the stream function (\ref{eq:fourier_stream_function}) gives
\begin{linenomath*}
\begin{equation} \label{eq:fourier_tangential}
2AI_{1}\left(K\right) + 2BK_{1}\left(K\right) + \tilde{C}\left[I_{0}\left(K\right) + I_{2}\left(K\right)\right] + \tilde{D}\left[K_{0}\left(K\right) + K_{2}\left(K\right)\right] = 0.
\end{equation}
\end{linenomath*}
Next, we substitute the expansion (\ref{eq:expansions}) into the normal stress condition (\ref{eq:normal}), replacing $\hat{p}_{k}$ with $\hat{\psi}_{k}$ by taking $\partial_{z}$ of equation (\ref{eq:normal}) and using equation (\hyperref[eq:axisymmetric-stokes]{6b}). We then replace $\hat{\xi}_{k}$ by $\hat{\psi}_{k}$ using the kinematic condition (\ref{eq:kinematic}). The leading order result is
\begin{linenomath*}
\begin{equation} \label{eq:fourier_normal}
AF_{A} + BF_{B} + \tilde{C}F_{\tilde{C}} + \tilde{D}F_{\tilde{D}} = 0,
\end{equation}
\end{linenomath*}
where
\begin{linenomath*}
\begin{subequations} \label{eq:F_coeffs}
\begin{eqnarray}
F_{A} &=& -3I_{0}\left(K\right)/2 + \left[1/K - K/\Sigma + 1/\Sigma K\right]I_{1}\left(K\right) - I_{2}\left(K\right) \\
F_{B} &=& 3K_{0}\left(K\right)/2 + \left[1/K - K/\Sigma + 1/\Sigma K\right]K_{1}\left(K\right) + K_{2}\left(K\right) \\
F_{\tilde{C}} &=& 2I_{0}\left(K\right)/K - 2I_{1}\left(K\right) + [2/K - \allowbreak K/\Sigma + 1/\Sigma K]\allowbreak I_{2}\left(K\right) \\
F_{\tilde{D}} &=& 2K_{0}\left(K\right)/K + 2K_{1}\left(K\right) + \left[2/K - K/\Sigma + 1/\Sigma K\right]K_{2}\left(K\right).
\end{eqnarray}
\end{subequations}
\end{linenomath*}
Equations (\ref{eq:fourier_no_slip}-\ref{eq:fourier_normal}) are a set of four linear equations for the constants $A$, $B$, $\tilde{C}$, and $\tilde{D}$. For there to be non-trivial solutions, we require that the matrix of the linear system has vanishing determinant. This condition establishes the dispersion relation between $\Sigma$ and $K$,
\begin{linenomath*}
\begin{equation} \label{eq:dispersion}
\Sigma\left(K;S\right) = \frac{\left(K^2 - 1\right)~G_{1}\left(K, S\right)}{2K^2~G_{2}\left(K, S\right)},
\end{equation}
\end{linenomath*}
where
\begin{linenomath*}
\begin{multline} \label{eq:G_1}
G_{1} = 1 + KS((-KSI_0^2(KS) + 2I_0(KS)I_1(KS) + KSI_1^2(KS))K_1^2(K) \\ +  I_1^2(K)(-KS K_0^2(KS) - 2K_0(KS)K_1(KS) + KSK_1^2(KS)) \\ + I_1(K)K_1(K)(-KS + I_1(KS)((2 + K^2S^2) K_0(KS) - 2KS K_1(KS)) \\ + I_0(KS)(-2KSK_0(KS) + (-2 + K^2S^2)K_1(KS))))
\end{multline}
and
\begin{multline} \label{eq:G_2}
G_{2} = 1 + 2/K^2 + S^2 + KSI_0^2(K) (KS K_0^2(KS) + 2K_0(KS)K_1(KS) - KSK_1^2(KS)) \\ + S/K(I_0^2(KS)(K^3S K_0^2(K) - KS(1 + K^2)K_1^2(K)) \\ + I_1^2(KS) (-K^3SK_0^2(K) + KS(1 + K^2)K_1^2(K)) \\ + I_1(K)I_1(KS)K_1(K)((2 + K^2(2 + S^2))K_0(KS) - 2(1 + K^2)KSK_1(KS)) \\ + I_0(KS)(I_1(KS)(-2K^2K_0^2(K) + 2(1 + K^2)K_1^2(K)) \\ + I_1(K)K_1(K)(-2(1 + K^2)KSK_0(KS) + (-2 + K^2(-2 + S^2))K_1(KS))) \\ + I_1(K)(-KSK_1(K) + (1 + K^2)I_1(K)(-KSK_0^2(KS) - 2K_0(KS) K_1(KS) + KSK_1^2(KS)))) \\ - 2I_0(K)K_0(K)(-1 + K^2S^2(I_1(KS)K_1(KS) + I_0(KS)K_2(KS)))
\end{multline}
\end{linenomath*}
The observed dimensionless wavelength $\Lambda_{\mathrm{\max}}$ corresponds to the maximal growth rate, i.e., $\Lambda_{\mathrm{\max}}\left(S\right)=2\pi/\allowbreak\mathrm{argmax}_{K}\Sigma\left(K; S\right)$, and it depends only on the geometric parameter $S$ (\hyperref[fig:3]{Fig.~3b}).

One might suspect that the rheology of protein condensates is not truly Newtonian, and that viscoelastic effects might play a role \cite{Jawerth2018}. However, it has been shown that such effects are subdominant on wavelength selection in this instability, even at high Weissenberg number, which measures the magnitude of elastic effects. Finite viscoelasticity only ``swells'' the droplets without affecting their peak-to-peak spacing, as well as slows down the instability time scale \cite{Boulogne2012}.

In order to contextualize our experimental measurements, we can estimate the capillary velocity $\gamma/\mu$ by noting that the typical time scale for droplet formation $T\sim r_{\mathrm{o}}\mu/\gamma$. Using $r_{\mathrm{o}}=\mathrm{O}(10)~\mathrm{nm}$ as the typical length scale (from \hyperref[fig:2]{Fig.~2b}) and using $T$ between $\mathrm{O}(10)~\mathrm{s}$ (TIRF conditions; \hyperref[fig:1]{Fig.~1a} and \hyperref[movieCaptions]{Supplementary Movie 1}) and $\mathrm{O}(10^3)~\mathrm{s}$ (AFM conditions; \hyperref[fig:s7]{Supplementary Fig.~7b}), we estimate a capillary velocity $\gamma/\mu$ between $\mathrm{O}(1)~\mathrm{nm/s}$ and $\mathrm{O}(10^{-2})~\mathrm{nm/s}$. 

\subsection{Nucleation and growth of the condensed film} \label{theory:1-2}

In \hyperref[theory:1-1]{Supplementary Information 1.1}, we assumed that the nucleation and growth of the condensed film on the microtubule occurred much faster than the subsequent Rayleigh-Plateau instability, and so we neglected it in our hydrodynamic model. This is indeed what we observe experimentally. Here, we explicitly consider the nucleation and growth dynamics, thereby elucidating the precise conditions where we expect the observed time scale separation to hold.

We consider a suspension of microtubules in a pool of soluble protein, whose concentration field we denote $c=c\left(\boldsymbol{x},t\right)$. In our experiments, we spike in protein at a concentration $c\left(\boldsymbol{x},t=0\right)=c_{0}$, where $c_{0}$ is large enough such that there is a thermodynamic driving force for the soluble protein to phase separate into a protein-rich condensed phase of concentration $c_{\mathrm{R}}$ and a protein-poor ``gas'' phase of concentration $c_{\mathrm{P}}$, where $c_{\mathrm{P}}<c_{0}<c_{\mathrm{R}}$. $c_{\mathrm{P}}$ and $c_{\mathrm{R}}$ are constrained by the lever rule and are in general functions of $c_0$ \cite{shell2015thermodynamics}. For simplicity we take each microtubule to be periodically spaced by a distance $2\bar{R}$, so that the number density of microtubules in solution is $n = 1/\pi\bar{R}^2l$ (\hyperref[fig:s6]{Supplementary Fig.~6a}). Given this geometry we can also assume radial symmetry throughout: $c=c\left(r,t\right)$ only.

At $t=0$, phase separation begins, and the microtubules offer sites for heterogeneous nucleation of the condensed phase. For simplicity we neglect any homogeneous nucleation of condensed protein droplets in the bulk. We assume that all of the initial soluble protein pool eventually contributes to the condensed film on the microtubules, effectively reaching $c_{\mathrm{P}}=0$, which is a nonessential but mathematically simplifying assumption in polymer phase separation \cite{shell2015thermodynamics,Berry2018a}. As a check, we compute the ratio of fluorescence intensities between condensed TPX2 on the microtubule versus background soluble TPX2 (\hyperref[fig:s2]{Supplementary Fig.~2a}), finding $c_{\mathrm{P}}/c_{\mathrm{R}}$ between $10^{-2}$ and $10^{-1}$ over all tested concentrations, consistent with our assumption. 

We consider the transport of soluble protein towards the condensed film to be diffusive, so that $\partial_{t}c=D\nabla^2c$ in the bulk, where $D$ is the diffusivity constant of the soluble protein. We denote the position of the film interface as $r=\xi\left(t\right)$, which will grow in time as soluble protein from the bulk diffuses onto the film. Hence we can write the global mass balance
\begin{linenomath*}
\begin{subequations} \label{eq:global-balance}
\begin{eqnarray}
\frac{\mathrm{d}}{\mathrm{d}t}\left(c_{\mathrm{R}}\left[\pi\xi^{2}l - \pi r_{\mathrm{i}}^2l\right]\right) &=& -\frac{\mathrm{d}}{\mathrm{d}t}\int_{\xi}^{\bar{R}}2\pi lr\mathrm{d}r\ c\left(r, t\right) \\
\implies \xi^2\left(t\right) &=& \frac{c_{0}}{c_{\mathrm{R}}}\bar{R}^2 + r_{\mathrm{i}}^2\left(1 - \frac{c_{0}}{c_{\mathrm{R}}}\right) - 2\int_{\xi}^{\bar{R}}r\mathrm{d}r\ \frac{c\left(r,t\right)}{c_{\mathrm{R}}},
\end{eqnarray}
\end{subequations}
\end{linenomath*}
where in going from equation (\hyperref[eq:global-balance]{18a}) to equation (\hyperref[eq:global-balance]{18b}) we integrate in time and apply the initial conditions $c\left(r,t=0\right)=c_{0}$ and $\xi\left(t=0\right)=r_{\mathrm{i}}$. By just considering equation (\hyperref[eq:global-balance]{18b}) we see that the final film thickness $h = \xi\left(t\rightarrow\infty\right) - r_{\mathrm{i}}$ is reached after depleting the condensable solute $c\left(r, t\rightarrow\infty\right)\rightarrow0$. Therefore
\begin{linenomath*}
\begin{equation} \label{eq:final_film}
r_{\mathrm{o}}:=r_{\mathrm{i}} + h = \sqrt{\frac{c_{0}}{c_{\mathrm{R}}}\bar{R}^2 + r_{\mathrm{i}}^2\left(1 - \frac{c_{0}}{c_{\mathrm{R}}}\right)},
\end{equation}
\end{linenomath*}
where we have defined the final position of the film interface as $r_{\mathrm{o}}$ for later convenience. The main feature of (\ref{eq:final_film}) is that it predicts the film thickness $h$ to be an increasing function of protein concentration $c_0$ and a decreasing function of microtubule density $n$. To check if the relationship $h=h\left(c_{0}\right)$ as specified by equation (\ref{eq:final_film}) is consistent with our experimental AFM observations, we plot measured values of $h$ against $c_{\mathrm{0}}$ and fit the data to equation (\ref{eq:final_film}) (\hyperref[fig:s6]{Supplementary Fig.~6b}). In doing so we neglect the dependence of $c_{\mathrm{R}}$ on $c_0$. We observe qualitative agreement, suggesting that the final film thickness is indeed set by depletion of locally condensable solute.

At $r=\xi\left(t\right)$ there is a sink of soluble protein as it is adsorbed onto the condensed film, and so we demand that $c\left(r=\xi,t\right) = 0$, which enforces the driving force for the soluble protein to diffuse towards the film. Due to our assumption of a periodic array of microtubules, we can additionally enforce the symmetry condition $\partial_{r}c\left(r=\bar{R},t\right)=0$. This closes the problem statement for the desired fields $c$ and $\xi$, with the goal being to determine the dynamics of the film thickness $h\left(t\right) = \xi\left(t\right) - r_{\mathrm{i}}$.

To solve for the fields, we work with the rescaled variables $C=c/c_{0}$, $R = r/\bar{R}$, $I=\xi/r_{\mathrm{o}}$, and $T = t/\left(\bar{R}^2/D\right)$. The rescaling has been carefully chosen such that $C\in\left[0,1\right]$ and $I\in\left[S,1\right]$ for all parameter values. This choice dramatically speeds up the subsequent root finding calculations. Under these transformations the boundary value problem becomes
\begin{linenomath*}
\begin{subequations} \label{eq:bvp}
\begin{eqnarray}
\partial_{T} C &=& 1/R~\partial_{R}\left(R~\partial_{R} C\right) \\
C\left(R, T=0\right) &=& 1 \\
C\left(R=MI\left(T\right), T\right) &=& 0 \\
\partial_{R} C\left(R=1,T\right) &=& 0 \\
I\left(0\right) &=& S \\
I^2\left(T\right) &=& 1-\frac{2\beta}{M^2}\int_{MI\left(T\right)}^{1}R\mathrm{d}R\ C\left(R, T\right),
\end{eqnarray}
\end{subequations}
\end{linenomath*}
where $\beta = c_{0}/c_{\mathrm{R}}$ is a thermodynamic parameter, $S=r_{\mathrm{i}}/r_{\mathrm{o}}$ is a geometric ratio, and $M=r_{\mathrm{o}}/\bar{R}$ is a measure of the microtubule density. For the remainder of this section we take $M=1/10$. However, equation (\ref{eq:final_film}) relates these parameters, so only two are independent. Squaring equation (\ref{eq:final_film}), dividing through by $\bar{R}^2$, and using the definitions for $\beta$, $S$, and $M$ gives, after rearrangement,
\begin{linenomath*}
\begin{equation} \label{eq:parameter_constraint}
\beta = \frac{M^2\left(1-S^2\right)}{1-M^2S^2}.
\end{equation}
\end{linenomath*}
In our AFM experiments described in the main text, we directly measure $S$ and fix $M$, and so the only value of $\beta$ self-consistent with our theory is given by equation (\ref{eq:parameter_constraint}). For our experimentally observed range of $S \in \left[0.5, 0.7\right]$, we find $\beta = \mathrm{O}\left(10^{-2}\right)$, which is consistent with $c_{\mathrm{R}} > c_{0}$.

Equations (\ref{eq:bvp}) would be a standard boundary value problem were it not for the integral constraint (\hyperref[eq:bvp]{20f}). Nevertheless, we can proceed with the separation anzats $C=f_{T}\left(T\right)\phi\left(R; I\right)$ where the notation $\phi\left(R; I\right)$ signifies that we treat $I$ as a fixed parameter when treating $\phi$. This construction is valid so long as the dynamics of $I\left(T\right)$ quasi-statically adjusts to the dynamics of $C\left(R, T\right)$. Proceeding with the usual steps \cite{Greenberg2013}, the two boundary conditions (\hyperref[eq:bvp]{20c},\hyperref[eq:bvp]{d}) can be satisfied by the eigenfunctions $\phi = \phi_{n}\left(R\right)=J_{0}\left(\lambda_{n}R\right) - \left[J_{1}\left(\lambda_{n}\right)/Y_{1}\left(\lambda_{n}\right)\right]Y_{0}\left(\lambda_{n}R\right)$, where the eigenvalues satisfy the \textit{time-dependent} relation
\begin{linenomath*}
\begin{equation} \label{eq:eigenvalues}
J_{0}\left(\lambda_{n}MI\right)Y_{1}\left(\lambda_{n}\right) - Y_{0}\left(\lambda_{n}MI\right)J_{1}\left(\lambda_{n}\right) = 0.
\end{equation}
\end{linenomath*}
$J_{p}$ and $Y_{p}$ are Bessel functions of the first and second kind of order $p$, respectively. This is to say that for every $I=I\left(T\right)$ there is a corresponding set of eigenvalues $\lambda_n$. We also note that, for any choice of $I$, $\lambda_1<\lambda_2<\ldots$. We find that $f_{T}=e^{-\lambda_n^2T}$, so that the solution may be represented by
\begin{linenomath*}
\begin{equation} \label{eq:solution_bvp}
C\left(R, T\right) = \sum_{n=1}^{\infty}a_{n}e^{-\lambda_{n}^2T}\phi_{n}\left(R\right),
\end{equation}
\end{linenomath*}
where the $a_{n}$ are chosen to satisfy (\hyperref[eq:bvp]{20b}) and are calculated using orthogonality of the eigenfunctions under the inner product $\langle f, g\rangle = \int_{MI}^{1}R\mathrm{d}R\ fg$. We find $a_{n}=\langle \phi_{n}, 1\rangle/\langle\phi_{n},\phi_{n}\rangle$, where
\begin{linenomath*}
\begin{multline} \label{eq:inner_phi_1}
\langle \phi_{n}, 1\rangle = \lbrace J_{1}\left(\lambda_{n}\right) - MIJ_{1}\left(\lambda_{n}MI\right) + J_{1}\left(\lambda_{n}\right)/Y_{1}\left(\lambda_{n}\right)\left[MIY_{1}\left(\lambda_{n}MI\right) - Y_{1}\left(\lambda_{n}\right)\right]\rbrace/\lambda_{n}
\end{multline}
\end{linenomath*}
\begin{linenomath*}
\begin{multline} \label{eq:inner_phi_phi}
\langle \phi_{n}, \phi_{n}\rangle = \lbrace J_{0}^{2}(\lambda_{n}) + J_{1}^{2}(\lambda_{n}) - (MI)^{2}[J_{0}^{2}(\lambda_{n}MI) + J_{1}^{2}(\lambda_{n}MI)] \\ + 2J_{1}(\lambda_{n})/Y_{1}(\lambda_{n})[(MI)^2[J_{0}(\lambda_{n}MI)Y_{0}(\lambda_{n}MI) + J_{1}(\lambda_{n}MI)Y_{1}(\lambda_{n}MI)] \\ - J_{0}(\lambda_{n})Y_{0}(\lambda_{n}) - J_{1}(\lambda_{n})Y_{1}(\lambda_{n})]  + J_{1}^2(\lambda_{n})/Y_{1}^2(\lambda_{n})[Y_{0}^{2}(\lambda_{n}) + Y_{1}^{2}(\lambda_{n}) \\ - (MI)^{2}[Y_{0}^{2}(\lambda_{n}MI) + Y_{1}^{2}(\lambda_{n}MI)]] \rbrace/2.
\end{multline}
\end{linenomath*}
Upon substitution of the expansion (\ref{eq:solution_bvp}), the integral constraint (\hyperref[eq:bvp]{20f}) becomes
\begin{linenomath*}
\begin{equation} \label{eq:objective}
F\left(I\right) := I^2\left(T\right) + \frac{2\beta}{M^2}\sum_{n=1}^{\infty}a_{n}e^{-\lambda_{n}^2T}\langle\phi_{n}, 1\rangle - 1 = 0.
\end{equation}
\end{linenomath*}

The solutions $C\left(R, T\right)$ and $I\left(T\right)$ are constructed in the following self-consistent way. For every time $T$ we numerically compute the zero of $F\left(I\right)$ defined by equation (\ref{eq:objective}). This gives the value of $I$ that satisfies the integral constraint, which then fixes the values of $\lambda_{n}$ by equation (\ref{eq:eigenvalues}). Hence, the solution $C\left(R, T\right)$ is fully determined by the expansion (\ref{eq:solution_bvp}). We plot the position of the film interface $I\left(T\right)$ for various $S$ within our experimentally probed range (\hyperref[fig:s6]{Supplementary Fig.~6c}).

Equation (\ref{eq:objective}) is a nonlinear, implicit relation for $I\left(T\right)$, and therefore admits no succinct closed form solution. However, we can derive explicit asymptotic representations for $I\left(T\right)$ working in the early and late time limits.

For early times $t\rightarrow0$, the soluble bulk has hardly been depleted, so it is as if the film grows in an unbounded medium of solute. Therefore we may replace the integral constraint (\hyperref[eq:bvp]{20f}) with
\begin{linenomath*}
\begin{subequations} \label{eq:flux_condition}
\begin{eqnarray}
\frac{\mathrm{d}}{\mathrm{d}t}\left(c_{\mathrm{R}}\left[\pi\xi^{2}l - \pi r_{\mathrm{i}}^2l\right]\right) &=& \left(D\frac{\partial c}{\partial r}\bigg{|}_{r=\xi}\right)2\pi\xi l\\
\implies \frac{\mathrm{d}\xi}{\mathrm{d}t} &=& \frac{D}{c_{\mathrm{R}}}\frac{\partial c}{\partial r}\bigg{|}_{r=\xi},
\end{eqnarray}
\end{subequations}
\end{linenomath*}
which is the statement that the film grows due to the diffusive flux of solute at $r=\xi\left(t\right)$. Moreover, the film interface is far from reaching its final value $r_{\mathrm{o}}$, so to analyze this limit we choose the different rescaling $X=r/r_{\mathrm{i}} - 1$, $\delta=\xi/r_{\mathrm{i}} - 1$, and $\tilde{T}=t/\left(r_{\mathrm{i}}^2/D\right)$. This rescaling measures distances from the surface of the microtuble, so that $\delta\ll1$ for $\tilde{T}\rightarrow0$. Hence the boundary value problem appropriate for early times becomes
\begin{linenomath*}
\begin{subequations} \label{eq:early-bvp}
\begin{eqnarray}
\partial_{\tilde{T}} C &=& \partial_{X}^2 C + \partial_{X} C / \left(1+X\right) \\
C\left(X, \tilde{T}=0\right) &=& 1 \\
C\left(X=\delta\left(\tilde{T}\right), \tilde{T}\right) &=& 0 \\
C\left(X\rightarrow\infty,\tilde{T}\right) &\rightarrow& 1 \\
\delta\left(0\right) &=& 0 \\
\frac{\mathrm{d}\delta}{\mathrm{d}\tilde{T}} &=& \beta\frac{\partial C}{\partial X}\bigg{|}_{X=\delta}.
\end{eqnarray}
\end{subequations}
\end{linenomath*}

We now introduce the similarity transformation $\eta=X/\delta$ and demand that $C=C\left(\eta\right)$ only. Under these conditions equation (\hyperref[eq:early-bvp]{28a}) becomes
\begin{linenomath*}
\begin{equation} \label{eq:similarity_transformed_1}
-\eta\frac{1}{\delta}\frac{\mathrm{d}\delta}{\mathrm{d}\tilde{T}}C' = \frac{C''}{\delta^2} + \frac{C'}{\delta\left(1 + \eta\delta\right)},
\end{equation}
\end{linenomath*}
where $(.)'=\mathrm{d}(.)/\mathrm{d}\eta$. For $\tilde{T}\rightarrow0$, $\eta=\mathrm{O}\left(1\right)$ near the film and $\delta\ll1$, hence we can neglect $\eta\delta$ compared to $1$. Doing so and rearranging gives
\begin{linenomath*}
\begin{equation} \label{eq:similarity_transformed_2}
-\left(\eta\delta\frac{\mathrm{d}\delta}{\mathrm{d}\tilde{T}} + \delta\right)C' = C''.
\end{equation}
\end{linenomath*}
On physical grounds we assume the ordering $\mathrm{O}\left(\mathrm{d}\delta/\mathrm{d}\tilde{T}\right) > \mathrm{O}\left(\delta\right)$, so we can again ignore the isolated $\delta$ to find
\begin{linenomath*}
\begin{equation} \label{eq:similarity_transformed_3}
-\eta\left(\delta\frac{\mathrm{d}\delta}{\mathrm{d}\tilde{T}}\right)C' = C''.
\end{equation}
\end{linenomath*}
To satisfy $C=C\left(\eta\right)$ only, we are forced to choose $\delta\mathrm{d}\delta/\mathrm{d}\tilde{T} = b$ for some constant $b$ to be determined self-consistently later. Using the initial condition (\hyperref[eq:early-bvp]{28e}) this equation integrates to $\delta = \sqrt{2b\tilde{T}}$, which allows for the three conditions (\hyperref[eq:early-bvp]{28b}-\hyperref[eq:early-bvp]{d}) to collapse to only two conditions $C\left(\eta=1\right)=0$ and $C(\eta\rightarrow\infty)\rightarrow1$. Using these conditions the solution to equation (\ref{eq:similarity_transformed_3}) is
\begin{linenomath*}
\begin{equation} \label{eq:similarity_solution}
C = 1 - \frac{\mathrm{erfc}\left(\eta/\sqrt{2}\right)}{\mathrm{erfc}\left(1/\sqrt{2}\right)},
\end{equation}
\end{linenomath*}
where $\mathrm{erfc}\left(z\right)=2/\sqrt{\pi}\int_{z}^{\infty}\mathrm{d}x~e^{-x^2}$ is the complementary error function. Substituting equation (\ref{eq:similarity_solution}) into equation (\hyperref[eq:early-bvp]{28f}) determines $b=\delta\mathrm{d}\delta/\mathrm{d}\tilde{T}=2\beta/\left(\sqrt{2\pi e}~\mathrm{erfc}\left(1/\sqrt{2}\right)\right)$. The final asymptotic formula for the film thickness is therefore
\begin{linenomath*}
\begin{equation} \label{eq:early_time_solution}
\frac{\xi}{r_{\mathrm{i}}} \sim 1 + \sqrt{\frac{4\beta}{\sqrt{2\pi e}~\mathrm{erfc}\left(1/\sqrt{2}\right)}\tilde{T}}~\mathrm{for}~\tilde{T}\rightarrow0,
\end{equation}
\end{linenomath*}
which displays the typical $\sim\sqrt{\tilde{T}}$ behavior expected for an interface that grows by diffusive processes.

For late times $t\rightarrow\infty$, the leading term $n=1$ in the expansion (\ref{eq:solution_bvp}) surely dominates. Moreover, $I\left(T\rightarrow\infty\right)\sim1$, so we can simply evaluate the first eigenvalue $\lambda_1$ using equation (\ref{eq:eigenvalues}) taking $I=1$. We find $\lambda_1\simeq1.103$, and equations (\ref{eq:solution_bvp}) and (\ref{eq:objective}) simplify to
\begin{linenomath*}
\begin{subequations} \label{eq:late-time-solution}
\begin{eqnarray}
C\left(R, T\right) &\sim& \left(a_1\phi_{1}\left(R\right)\right)\big{|}_{I=1}e^{-\lambda_1^2T} \\
I\left(T\right) &\sim& \sqrt{1 - 2\beta/M^2\left(a_1\langle\phi_1,1\rangle\right)\big{|}_{I=1}e^{-\lambda_1^2T}}~\mathrm{for}~T\rightarrow\infty.
\end{eqnarray}
\end{subequations}
\end{linenomath*}
Fortuitously, equation (\hyperref[eq:late-time-solution]{34b}) gives a uniformly excellent approximation to the exact result for all times (\hyperref[fig:s6]{Supplementary Fig.~6c}).

In this section, we have established the dynamics for the nucleation and growth of a condensed film on a microtubule. As long as these dynamics are fast compared to the subsequent Rayleigh-Plateau instability, i.e. $\bar{R}^2/D \ll r_{\mathrm{o}}\mu/\gamma$, we may neglect them in the hydrodynamic analysis. Substituting the estimates $r_{\mathrm{o}}\mu/\gamma\simeq1/0.03~\mathrm{min}$ (from \hyperref[fig:s7]{Supplementary Fig.~7b}), $\bar{R}\simeq1~\mu\mathrm{m}$, and $D\simeq10~\mu\mathrm{m}^2/\mathrm{s}$ we find that $\bar{R}^2/D=\mathrm{O}(10^{-1})~\mathrm{s}$ and $r_{\mathrm{o}}\mu/\gamma=\mathrm{O}(10^{3})~\mathrm{s}$. Hence time scale separation is well satisfied.

\subsection{Kinetic Monte Carlo simulations} \label{theory:1-3}

We now consider the stochastic chemical kinetics involved in $F$ branching factors binding to and unbinding from a microtubule lattice of length $l$ coated with condensed protein. We wish to determine the first time $\tau$ at which $F$ factors find each other on the protein coated lattice for two different coating geometries: a uniform coating and a periodic coating of droplets with spacing $\lambda$. This serves as a model for how multiple necessary factors may find each other on a microtubule and subsequently nucleate a branch, and we seek to answer the question of whether a periodic coating of protein droplets accelerates this process.

For concreteness, we choose $l=5~\mu\mathrm{m}$ and $\lambda=0.25~\mu\mathrm{m}$, although these choices do not affect the results. We fix $d=25~\mathrm{nm}$ to be the distance below which we consider the $F$ factors to have colocalized, which is the typical size of a $\gamma$-TuRC. Let $N^{\left(u\right)}_{\alpha}\left(t\right)$ and $N^{\left(b\right)}_{\alpha}\left(t\right)$ be the random variables specifying the number of well-mixed unbound and bound factors of species $\alpha$, respectively, where $\alpha=1,2,\ldots,F$. We start the dynamics with $N^{\left(u\right)}_{\alpha}\left(0\right) = N$ and $N^{\left(b\right)}_{\alpha}\left(0\right) = 0$. Each distinct factor can bind to the coated lattice with rate constant $k_{\mathrm{on}}$ and unbind with rate constant $k_{\mathrm{off}}$. Factors can only bind to the coated lattice (\hyperref[fig:4]{Fig.~4a}), so the total length of available binding sites is $L=l$ for a uniform coating and $L=l/2$ for a periodic coating. One dimensionless number we anticipate affects the dynamics is the ratio of unbinding to binding timescales, which we define as $s=\frac{1/k_{\mathrm{off}}}{1/k_{\mathrm{on}}l}$. We set $k_{\mathrm{on}}=1.1\cdot10^{-3}~\mu\mathrm{m}~\mathrm{sec}^{-1}$ \cite{Thawani2019} and study the dynamics with respect to $s$, keeping in mind that microtubule associated proteins bind tightly so that $s>1$. We assume binding and unbinding events are Poisson processes with rates $r_{\alpha,\mathrm{on}}\left(t\right)=k_{\mathrm{on}}LN^{\left(u\right)}_{\alpha}\left(t\right)$ and $r_{\alpha,\mathrm{off}}\left(t\right)=k_{\mathrm{off}}N^{\left(b\right)}_{\alpha}\left(t\right)$, respectively. 

The dynamics are updated using a standard Gillespie algorithm \cite{Gillespie1977}. In this scheme time is updated by the rule $t_{i+1} = t_{i} - \log\left(1 - u\right)/m$, where $u$ is a uniform random variable on $\left[0,1\right]$ and $m = \sum_{\alpha}r_{\alpha,\mathrm{on}}\left(t_i\right) + r_{\alpha,\mathrm{off}}\left(t_i\right)$. One event is then realized at time $t_{i+1}$, a binding event of a factor of species $\alpha$ with probability $P^{\left(b\right)}_{\alpha}=r_{\alpha,\mathrm{on}}\left(t_i\right)/m$ or an unbinding event of a factor of species $\alpha$ with probability $P^{\left(u\right)}_{\alpha}=r_{\alpha,\mathrm{off}}\left(t_i\right)/m$. If a binding event occurs, the factor may bind anywhere on the coated lattice with uniform probability. The dynamics are stopped once any $F$ distinct factors come within a distance $d$ of each other, and the colocalization time $\tau$ is reported. 

Considering $F=2$, which is the minimum number of distinct factors needed for branching nucleation, we find that a periodic coating is more efficient at colocalizing factors than a uniform coating for all relevant values of $s$ and $N$ (\hyperref[fig:s9]{Supplementary Fig.~9a}). Of course, as $N$ increases the time $\tau$ to colocalize decreases, as there are simply more factors that can sample the lattice. As $s$ decreases, the factors are more likely to unbind and hence can more efficiently sample the lattice, and so $\tau$ correspondingly decreases. We also provide a typical histogram that shows a long time tail, characteristic of Poissonian chemical kinetics (\hyperref[fig:s9]{Supplementary Fig.~9b}).

\section{Experimental materials and methods} \label{methodsAll}

\subsection{Protein expression and purification}

Recombinant GFP-TPX2 was purified as previously described~\cite{Thawani2019, Alfaro-Aco2017}.
Full-length TPX2 or C-terminal TPX2 (amino acids 377 to 716) tagged on the N-terminus with Strep-6xHis-GFP-TEV was cloned into the pST50 expression vector.
TPX2 was expressed in E.~coli (strain Rosetta 2) for 7 hours at 25$^{\circ}$C and the cells lysed using EmulsiFlex (Avestin) in lysis buffer (50 mM tris-HCl, pH 8.0, 150mM imidazole, 750 mM NaCl, 2.5 mM PMSF, 6 mM BME, 1 cOmplete\texttrademark~EDTA-free Protease Inhibitor tablet (Sigma), and 1000 U DNase I (Sigma)).
Then, the protein was affinity purified using Ni-NTA beads in binding buffer (50 mM tris-HCl pH 8.0, 750 mM NaCl, 15 mM imidazole, 2.5 mM PMSF, 6 mM BME).
Next, the protein was eluted with 200 mM imidazole, and the protein was pooled and diluted 4-fold to a final NaCl concentration of 200 mM.
After, nucleotides were removed from the eluted protein with a HiTrap Heparin HP column (GE Healthcare).
Protein was bound to the column in 250 mM NaCl and then isocratic elution was done in 750 mM NaCl.
Solutions were prepared in Heparin buffer (50 mM tris-HCl, pH 8.0, 2.5 mM PMSF, 6 mM BME).
Finally, peak fractions were pooled and loaded onto a Superdex 200 pg 16/600 gel filtration column. 
Gel filtration was performed in CSF-XB buffer (10 mM K-HEPES, pH 7.7, 100 mM KCl, 1 mM MgCl$_2$, 5 mM EGTA) with 10\% (w/v) sucrose.
TPX2 was flash frozen in liquid nitrogen and stored at -80$^{\circ}$C until use. This TPX2 was used for all fluorescence, electron, and atomic force microscopy experiments.

Recombinant GFP-tagged augmin holocomplex was purified as previously described~\cite{Alfaro-Aco2019,song2018mechanism}.
Briefly, Sf9 cells were coinfected with several baculoviruses, each carrying a subunit of the augmin complex, at multiplicity of infection (MOI) values of 1--3.
Cells were collected and lysed 72 hours after infection.
Augmin subunit HAUS6 had an N-terminal ZZ-tag, subunit HAUS2 had a C-terminal GFP-6xHis, and the other six subunits were untagged.
The complex was affinity purified using IgG-Sepharose (GE Healthcare) and eluted via cleavage with 100–200 $\mu$g of GST-HRV3C protease.
The HRV3C protease was subsequently removed using a GSTrap 5 mL column (GE Healthcare).
The sample was further purified and concentrated using Ni-NTA agarose beads.
The protein complex was dialyzed overnight into CSF-XB with 10\% (w/v) sucrose.
Augmin was flash frozen in liquid nitrogen and stored at -80$^{\circ}$C until use.

$\gamma$-TuRC was purified from \textit{X. laevis} meiotic cytoplasm as done previously~\cite{Alfaro-Aco2019}.
First, 5-10 mL of cytoplasm was prepared from \textit{X. leavis} eggs using standard procedures~\cite{murray1989cyclin,hannak2006investigating} and 10-fold diluted in CSF-XB with 10\% sucrose (w/v), 10 $\mu$g each of leupeptin, pepstatin, and chymostatin, 1 mM GTP, and 1 mM DTT.
The larger particles in the cytosol were removed by low-speed centrifugation at 3000 g for 10 minutes at 4$^{\circ}$C.
The supernatant was 2-fold diluted with buffer and passed through 1.2 $\mu$m, 0.8 $\mu$m and 0.22 $\mu$m pore size filters.
$\gamma$-TuRC was precipitated by adding a solution of 6.5\% (w/v) PEG 8000 to the filtrate on ice for 30 minutes. 
After centrifuging at 17000 g for 20 minutes at 4$^{\circ}$C, the pellet was resuspended in 15 mL of CSF-XB with 10\% sucrose (w/v), 10 $\mu$g each of leupeptin, pepstatin, and chymostatin, 1 mM GTP, 1 mM DTT, and 0.05\% (v/v) NP-40 detergent.
The resuspension was centrifuged at 136000 g for 7 minutes at 4$^{\circ}$C.
Then, protein A sepharose beads (GE Healthcare) were used to pre-clear the supernatant for 20 minutes at 4$^{\circ}$C.
The beads were removed via centrifugation.
2-4 mL of $\gamma$-tubulin antibody (Sigma) at 1
mg/mL was added and the mixture was rotated for 2 hours at 4$^{\circ}$C.
After this, 1 mL of washed protein A sepharose beads was incubated with the sample on a rotator for 2 hours at 4$^{\circ}$C.
The beads were collected by spinning and then transferred to a column with CSF-XB with 10\% sucrose (w/v), 10 $\mu$g each of leupeptin, pepstatin, and chymostatin, 1 mM GTP, 1 mM DTT, and 0.05\% (v/v) NP-40 detergent.
The beads were washed with CSF-XB with 10\% sucrose (w/v), 10 $\mu$g each of leupeptin, pepstatin, and chymostatin, 1 mM GTP, 1 mM DTT, and 150 mM additional KCl, then with CSF-XB with 1 mM ATP, and finally with CSF-XB to remove the ATP. 
For biotinylation of $\gamma$-TuRC, done for visualizing the reconstitution using fluorescence microscopy, 25 $\mu$M of NHS-PEG4-biotin (Thermo Scientific) was incubated with the beads in CSF-XB for 1 hour at 4$^{\circ}$C, and unreacted reagent was washed away with CSF-XB before elution.
To elute $\gamma$-TuRC from the beads, 2 mL of a $\gamma$-tubulin peptide (amino acids 412–451) at 0.5 mg/mL in CSF-XB was added to the column and incubated overnight. The eluted sample was collected from the column the following day.
$\gamma$-TuRC was then concentrated using a 100 kDa centrifuge filter (Amicon).
This concentrated sample was loaded onto a 10-50\% (w/w) continuous sucrose gradient in CSF-XB buffer with 10 $\mu$g each of leupeptin, pepstatin, and chymostatin, 1 mM GTP, and 1 mM DTT.
The gradient was centrifuged for 3 hours at 4$^{\circ}$C at 200000 g in a TLS55 rotor (Beckman).
Fractions of the sucrose gradient were collected manually from the top of the gradient.
The fractions with the highest $\gamma$-tubulin signal as determined by western blot were combined and concentrated as above using a 100 kDa centrifuge filter.
Purified $\gamma$-TuRC was used within two days without freezing and kept on ice.

The kinesin-1 fragment K560-GFP was purified as previously described~\cite{case1997directional}.
Amino acids 1-560 of human conventional kinesin tagged on the C-terminus with GFP and 6xHis was cloned into pET17B vector~\cite{case1997directional}.
K560-GFP was expressed in E.~coli for 4 hours at 37$^{\circ}$C and the cells lysed as above using lysis buffer (50 mM NaPO$_4$, pH 8.0, 250 mM NaCl, 1 mM MgCl$_2$, 20 mM imidazole, 10 mM BME, 0.5 mM PMSF, 0.5 mM ATP, 1 cOmplete\texttrademark~EDTA-free Protease Inhibitor tablet, 1000 U DNase I).
Then, K560-GFP was affinity purified using a HisTrap HP column (GE Healthcare) using binding buffer (50 mM NaPO$_4$, pH 8.0, 250 mM NaCl, 1 mM MgCl$_2$, 20 mM imidazole, 10 mM BME, 0.5 mM PMSF, 0.5 mM ATP) and eluted using binding buffer at 500 mM imidazole.
Peak fractions were pooled and dialyzed overnight at 4$^{\circ}$C into CSF-XB buffer.

Bovine brain tubulin was labelled with Cy5 or Alexa568 NHS ester (GE Healthcare) or biotin-PEG4-NHS (Thermo Scientific) using methods previously described ~\cite{gell2010microtubule}.

Protein concentrations were determined by SDS-PAGE and then Coomassie staining of a concentration series of the protein of interest alongside a BSA standard, or using Bradford dye (Bio-Rad).

\subsection{Expression and purification of RanQ69L and EB1-mCherry for \textit{Xenopus laevis} meiotic cytosol experiments}

Proteins used to visualize branching microtubule nucleation in \textit{X.~laevis} meiotic cytosol, RanQ69L and EB1-mCherry, were purified as previously described~\cite{king2016visualizing}. RanQ69L, which was N-terminally BFP-tagged to improve solubility, was expressed and lysed as above in lysis buffer (100 mM tris-HCl, pH 8.0, 450 mM NaCl, 1 mM MgCl$_2$, 1 mM EDTA, 0.5 mM PMSF, 6 mM BME, 200 $\mu$M GTP, 1 cOmplete\texttrademark~EDTA-free Protease Inhibitor, 1000 U DNAse I). RanQ69L was affinity purified using a StrepTrap HP 5 mL column (GE Healthcare) in binding buffer (100 mM tris-HCl, pH 8.0, 450 mM NaCl, 1 mM MgCl$_2$, 1 mM EDTA, 0.5 mM PMSF, 6 mM BME, 200 $\mu$M GTP) and eluted using binding buffer with 2.5 mM D-desthiobiotin.
Eluted protein was dialyzed overnight into CSF-XB with 10\% (w/v) sucrose.
EB1-mCherry was expressed and lysed as above in lysis buffer (50 mM NaPO$_4$, pH 7.4, 500 mM NaCl, 20 mM imidazole, 2.5 mM PMSF, 6 mM BME, 1 cOmplete\texttrademark~EDTA-free Protease Inhibitor, 1000 U DNAse I), affinity purified using a HisTrap HP 5 mL column (GE Healthcare) in binding buffer (50 mM NaPO$_4$, pH 7.4, 500 mM NaCl, 20 mM imidazole, 2.5 mM PMSF, 6 mM BME) and eluted using binding buffer with 500 mM imidazole.
Finally, peak fractions were pooled and loaded onto a Superdex 200 pg 16/600 gel filtration column. 
Gel filtration was performed in CSF-XB with 10\% (w/v) sucrose.

\subsection{\textit{Xenopus laevis} meiotic cytosol preparation and immunodepletion}

Meiotic cytosol, also known as CSF extract or egg extract, was prepared from \textit{Xenopus laevis} eggs using standard protocols~\cite{murray1989cyclin,hannak2006investigating}.
Cytosol was diluted up to 75\% for TIRF microscopy experiments.
For depletion of TPX2 from cytosol, 30 $\mu$g of purified anti-TPX2 antibody was coupled to 200 $\mu$L protein A magnetic beads (Dynabeads, ThermoFisher) overnight.
Then, 200 $\mu$L of fresh meiotic cytosol was depleted in two rounds by incubating with 100 $\mu$L beads for 30 minutes in each round.
Control depletion was done using rabbit IgG antibody (Sigma). The efficiency of depletion was determined by western blot and confirmed using functional tests.
\textit{Xenopus laevis} frogs were cared for in accordance with the recommendations in the Guide for the Care and Use of Laboratory Animals of the National Institutes of Health. All animals were cared for according to the approved Institutional Animal Care and Use Committee (IACUC) protocol 1941-16 of Princeton University.

\subsection{Polymerization of GMPCPP-stabilized microtubules}

For TIRF microscopy experiments, double-cycled GMPCPP-stabilized microtubules were prepared~\cite{gell2010microtubule}.
Bovine tubulin at 20 $\mu$M was polymerized in BRB80 (80 mM K-PIPES, pH 6.8, 1 mM EGTA, 1 mM MgCl$_2$), with 10 mM GMPCPP (Jena Bioscience), 2 mM Alexa568-labelled tubulin, and 2 mM biotin-labelled tubulin, at 37$^{\circ}$C for 30 minutes.
Microtubules were pelleted via centrifugation at 126,000\ $g$ for 8 minutes, and depolymerized by resuspending and incubating in ice cold BRB80 totalling 80\% of original volume for 20 minutes.
GMPCPP was then added up to 10 mM, and microtubules were again polymerized and pelleted as in the first cycle.
Microtubules were flash frozen in liquid nitrogen and quickly thawed prior to use.
Microtubules were used at a final 1000-fold to 2000-fold dilution for TIRF experiments.

For electron and atomic force microscopy, bovine tubulin at 20 $\mu$M was incubated in BRB80 with 10 mM GMPCPP (Jena Bioscience) on ice for 5 minutes and then centrifuged 90 kRPM in a TLA100 rotor (Beckman) at 4$^{\circ}$C for 15 minutes.
The supernatant was then incubated at 37$^{\circ}$C for 30 to 60 minutes to polymerize microtubules.
Finally, the reaction mixture was 5-fold diluted and microtubules were pelleted via centrifugation as above at 27$^{\circ}$C for 5 minutes.
The supernatant was discarded and microtubules resuspended in the original volume of BRB80.
Microtubules were made fresh for each imaging day and diluted 500-fold for electron microscopy and 10-fold for atomic force microscopy.

\subsection{Reconstitution of microtubule branching nucleation \textit{in vitro}} \label{methods:reconstitution}

Branching microtubule nucleation was reconstituted \textit{in vitro} using purified proteins as was done recently~\cite{Alfaro-Aco2019}.
Briefly, microtubules were attached to a flow chamber made with PEG-passivated and biotinylated glass coverslips.
A mixture of TPX2 (0.05 $\mu$M), augmin (0.05 $\mu$M), and $\gamma$-TuRC was incubated on ice for 5 minutes and then flowed into the chamber.
Proteins were allowed to bind microtubules and excess protein was flowed out.
Then, BRB80 with 30 mM KCl, 1 mM GTP, 5 mM BME, 0.075\% (w/v) methylcellulose (4000 cP), 1\% (w/v) glucose, 0.02\% (v/v) Brij-35, 250 nM glucose oxidase, 64 nM catalase, 1 mg/mL BSA, 19 $\mu$M unlabeled bovine tubulin and 1 $\mu$M Cy5-labeled bovine tubulin was flowed into the chamber to 
initiate the nucleation and growth of branched microtubules, which was visualized via TIRF microscopy.

\subsection{TIRF microscopy sample preparation and imaging} \label{methods:tirf}
Coverslip-bottomed culture well plates (Grace BioLabs) were functionalized with anti-Biotin antibody (Abcam) at 0.1 mg/mL concentration at room temperature for 10 minutes.
Excess antibody was washed out by serially diluting in buffer (CSF-XB at 50-60 mM KCl or BRB80) three times, and then the coverslip blocked with $\kappa$-casein at 1 mg/mL.
Excess $\kappa$-casein was washed out by serially diluting in buffer as above.
Then, microtubules were allowed to settle and bind the coverslip surface for 10 minutes at room temperature and then unbound microtubules were removed via serial dilution as above.
TPX2 was then pipetted into the wells during image acquisition, to visualize TPX2 coating and forming droplets on microtubules.

Total internal reflection fluorescence (TIRF) microscopy was done using a Nikon TiE microscope with a 100x magnification, 1.49 NA objective.
An Andor Zyla sCMOS camera was used for acquisition. Images were acquired using NIS-Elements software (Nikon).
For imaging of the reconstitution of branching microtubule nucleation, the objective was heated to 33$^{\circ}$C using an objective heater (Bioptechs). 

\subsection{TIRF data analysis} \label{methods:tirfAnalysis}

Droplet sizes and spacings used to create \hyperref[fig:s1]{Supplementary Fig.~1a} were calculated manually using FIJI (ImageJ) \cite{schindelin2012fiji}. To calculate the average power spectrum in \hyperref[fig:s1]{Supplementary Fig.~1b}, line scans of droplet patterned microtubules in the raw TIRF images were taken in FIJI and Fourier transformed in MATLAB. The power spectrum is given by the formula 
\begin{linenomath*}
\begin{equation} 
\label{eq:powerSpectrumImg}
P(f) = |\hat{I}(f)|^2 = \left|\frac{1}{L}\int_{0}^{L} I(z)e^{-2\pi if z}dz\right|^2,
\end{equation}
\end{linenomath*} 
where $I$ is the intensity, $z$ is the position along the microtubule, $L$ is the length of the microtubule, and $f$ is a spatial frequency. These power spectra were then averaged. The frequency at which the average power spectrum is maximal corresponds to the average observed droplet spacing, as is also the case for the AFM data (\hyperref[methods:afmDataAnalysis]{Methods}). 

\subsection{Visualizing branching microtubule nucleation from TPX2-coated microtubules in \textit{X. laevis} meiotic cytosol} \label{methods:branching}

Visualization of branching microtubule nucleation from TPX2-coated microtubules was done as previously described~\cite{Thawani2019}.
Briefly, coverslips were etched and cleaned using KOH and then silanized, following a previously described protocol~\cite{gell2010microtubule}, and used to make a microscope flow chamber.
Then, a TPX2 fragment (amino acids 320–631) that does not promote or inhibit branching microtubule nucleation~\cite{Alfaro-Aco2017} was flowed into the chamber at 4 $\mu$M.
The protein was allowed to bind the coverslip at room temperature for 10 minutes.
Excess protein was washed out using BRB80.
Then, $\kappa$-casein at 1 mg/mL was flowed in and incubated at room temperature for 10 minutes to passivate unfunctionalized regions of the coverslip, 
and washed out as above.
Next, TPX2-depleted meiotic cytosol with 0.9 $\mu$M Cy5-labeled tubulin and 0.5 mM vanadate was flowed into the chamber.
Microtubules generated in cytosol were allowed to bind the coverslip.
The cytosol was washed out using CSF-XB with 10\% sucrose.
0.1 $\mu$M GFP-TPX2 was incubated with meiotic cytosol for 20 minutes, and then this cytosol supplemented with 0.1 $\mu$M GFP-TPX2, 10 $\mu$M ranQ69L, 0.9 $\mu$M cy5-labeled tubulin, 0.2 $\mu$M EB1-mCherry and 0.5 mM vanadate was flowed into the chamber. Initial TPX2 coating of the pre-attached microtubules and subsequent branching nucleation was visualized using TIRF microscopy.

\subsection{Electron microscopy sample preparation and imaging} \label{methods:em}

For experiments with TPX2 alone, microtubules were incubated for 5 minutes at room temperature on plasma-treated carbon grids (Electron Microscopy Sciences). Excess microtubules were wicked away, and then TPX2 diluted in BRB80 was pipetted onto the grids.
TPX2 was incubated with microtubules on the grids for 1 hour to let droplets form, and kept in a humidity chamber to prevent evaporation.

For experiments to see colocalization of $\gamma$-TuRC with TPX2 droplets,
microtubules were first incubated with a mixture of 50 nM TPX2, 50 nM augmin, and $\gamma$-TuRC.
The sample was diluted 10-fold with BRB80 and 5 $\mu$L of the diluted sample was immediately applied onto plasma-treated carbon grids.
Then, for both experiments, all liquid was wicked away and samples were stained by blotting the grids three times with 2\% uranyl acetate, after which the grids were left to dry. 
Images were collected with a CM100 TEM (Philips) operated at 80 keV with 64000x magnification and recorded using an ORCA camera.

\subsection{AFM sample preparation and imaging} \label{methods:afm}

Microtubules were bound to mica disks using divalent cations, as previously described~\cite{hamon2010high}, as follows.
Mica disks were mounted on metal disks using double-sided sticky tape (supplies all from Ted Pella).
A fresh mica layer was exposed by peeling off the previous layer using tape, and 1 M MgCl$_2$ was pipetted onto the mica disk and incubated at room temperature for 10 minutes.
MgCl$_2$ was blown off using nitrogen gas and microtubules were immediately deposited onto the mica disk and allowed to bind for 10 minutes.
Unbound microtubules were washed off using BRB80.
The mica disk was mounted onto the AFM stage using a magnet.

AFM was done using the Bruker BioScope Resolve operated in peak force fluid tapping mode using PEAKFORCE-HIRS-F-A or -B probes (Bruker), with peak force set to 25-40 pN.
These probes have a nominal tip radius of 1 nm and a stiffness of 0.35 N/m or 0.12 N/m, respectively.
A 1 Hz scan rate per pixel row was used to sample a $2~\mu\mathrm{m} \times 2~\mu\mathrm{m}$ area every $\simeq$ 4.25 minutes.
Resulting topographies were $256 \times 256$ pixels, yielding a lateral pixel size of about 8 nm.
After uncoated microtubules were imaged, TPX2 or kinesin-1 was gently pipetted into the fluid meniscus above the mica disk and below the AFM scan head so that protein can bind microtubules while the AFM was scanning the sample.
The AFM continued scanning the sample for 40 to 180 minutes as TPX2 droplets formed.
We waited a similar amount of time to probe kinesin-bound microtubules.
For most experiments, once TPX2 droplets had formed, after about 40 minutes, we moved the stage to a different area of the sample to collect more data from microtubules that were not probed as the droplets formed.
This confirms that the act of poking TPX2-coated microtubules repetitively with the AFM probe does not generate the patterns.
For these microtubules, for which the initial film thickness was not measured, the film thickness used for \hyperref[fig:3]{Fig.~3c} was the average of all film thicknesses measured in that sample, which was always within $\simeq~$1--2 nm.
For the highest TPX2 concentration range tested, $0.8\pm0.4~\mu$M, larger topographies ($4~\mu\mathrm{m} \times 4~\mu\mathrm{m}$, sampled $512 \times 512$ at a 0.5 Hz scan rate per pixel row so that the speed of the AFM probe was the same across the sample as in the smaller images) were taken once the patterns formed to enable sampling larger distances between droplets.

\subsection{AFM data analysis} \label{methods:afmDataAnalysis}

Raw AFM topographies were processed using Gwyddion data visualization and analysis software \cite{nevcas2012gwyddion} to remove horizontal scarring and linear and polynomial background, and to align rows along the scan axis. Topographical line scans along bare, uncoated microtubules and along microtubules 
within one frame ($\simeq 4$ minutes) of
the initial TPX2 coating were drawn to determine the film thickness.
Successive topographies of the same area were registered in FIJI (ImageJ) \cite{schindelin2012fiji} using the StackReg and TurboReg packages \cite{thevenaz1998pyramid}.
To systematically measure the spacing between TPX2 droplets, power spectra of line scans along TPX2-droplet-coated microtubules were calculated using MATLAB.
Power spectra use the Fourier transform to decompose a line scan into a spectrum of the strength of all spatial frequencies in the original line scan.
Analogously to equation \ref{eq:powerSpectrumImg}, power spectra $P(f)$ were calculated by squaring the absolute value of the normalized finite Fourier transform $\hat{h}(f)$ of the height profiles $h(z)$, where $z$ is spatial position and $f$ is spatial frequency:
\begin{linenomath*}
\begin{equation} 
\label{eq:powerSpectrum}
P(f) = |\hat{h}(f)|^2 = \left|\frac{1}{L}\int_{0}^{L} h(z)e^{-2\pi if z}dz\right|^2.
\end{equation}
\end{linenomath*}
$L$ is the length of the height profile.
For each microtubule, multiple (at least 3) spectra were averaged.
The inverse of the frequency at which the peak power in the averaged spectrum occurs corresponds to the spacing between TPX2 droplets: $\lambda=1/f$, as is also the case for the fluorescence data (\hyperref[methods:tirfAnalysis]{Methods}).
Due to the noise in the AFM data, the size of the droplets were not measured, as they were manually measured for EM and TIRF microscopy data.
Microtubules that broke apart or were too short to sample low spatial frequencies, and sections of microtubules that curved or were under crisscrossing microtubules, were excluded from analysis. 
Some line scans were high-pass filtered via a large smoothing window ($\simeq~$300-400 nm) when they still exhibited local linear or polynomial background.
Line scans were never low-pass filtered for determining spectra.
To verify that the growth rate of the droplets was exponential in time, as predicted by theory, the power of the frequency corresponding to the droplet spacing was measured starting when TPX2 initially coated the microtubule to after the droplets had formed, for many microtubules.
The power was then plotted over time.

\subsection{AFM condensed phase concentration estimation} \label{methods:concentrationEstimation}

In our AFM experiments, we have as the volume of the entire reaction chamber, $V_{\mathrm{chamber}} \approx50~\mu\mathrm{L}$.
At a bulk TPX2 concentration $c_{\mathrm{bulk~TPX2}} = 100~\mathrm{nM}$, we have $N_{\mathrm{TPX2}}=c_{\mathrm{bulk~TPX2}}\cdot6.02\cdot10^{23}\cdot V_{\mathrm{chamber}}\approx 3 \cdot 10^9$ total molecules of TPX2 in the reaction chamber.

Based on the lengths of the microtubules measured in five AFM fields, we determine the microtubule length density $\rho_{\mathrm{MT~length}}$, or microtubule length per area of reaction chamber, to be $\rho_{\mathrm{MT~length}}=1.7 \cdot 10^{-3}\pm 9.4\cdot10^{-4}~\mathrm{nm}^{-1}$ (mean $\pm$ standard deviation). The reaction chamber rests on a mica desk with a diameter $d=12~\mathrm{mm}$. Thus, the total length of microtubules in the reaction chamber is $l_{\mathrm{total~MT}}=\rho_{\mathrm{MT~length}}\cdot A_{\mathrm{mica~disk}}\approx2\cdot 10^8~\mu\mathrm{m}$. We measured the initial thickness of the TPX2 film on the microtubule to be $h_{\mathrm{TPX2~film}}=13.4 \pm 3.1~\mathrm{nm}$ (mean $\pm$ standard error of the mean). Assuming an inner radius that is the radius of a microtubule, $r_i=12.5~\mathrm{nm}$, and using an outer radius $r_o=r_i+h_{\mathrm{TPX2~film}}$, this gives the following for the total volume of the TPX2 film on all microtubules in the reaction chamber, $V_{\mathrm{total~TPX2~film}}=\pi(r_o^2-r_i^2)l_{\mathrm{total~MT}}\approx3\cdot 10^5 \mu\mathrm{m}^3$.

Finally, assuming that all TPX2 molecules condense, we find the density of TPX2 in the condensed film to be $\rho_{\mathrm{TPX2~in~film}}=N_{\mathrm{TPX2}}/V_{\mathrm{total~TPX2~film}}\approx 10^4~\mu\mathrm{m}^{-3}$ or $\approx 16~\mu\mathrm{M}$. This estimate is consistent with equation (\ref{eq:parameter_constraint}) for $S=1/2$ and $M=1/10$. This puts the concentration of condensed TPX2 within the range of recent concentration measurements of condensed proteins \textit{in vivo} \cite{riback2020composition}.

\newpage

\section{Supplementary tables}

\begin{table}[h!] \label{table:s1}
\centering
\begin{tabularx}{\textwidth}{XXXXXX}
 Microtubule & \# of TPX2 droplets & \# of TPX2 droplets with $\geq$ 1 branch  & \# of TPX2 droplets with $\geq$ 2 branches & \# of TPX2 droplets with $\geq$ 1 branch
 
 /
  
 \# of TPX2 droplets & \# of TPX2 droplets with $\geq$ 2 branch
 
 /
  
 \# of TPX2 droplets  \\ \hline \hline
 1 & 10 & 6 & 1 & 0.60 & 0.10 \\ \hline
 2 & 7 & 4 & 2 & 0.57 & 0.29 \\ \hline
 3 & 4 & 4 & 3 & 1.0 &  0.75 \\ \hline
 4 & 8 & 6 & 2 & 0.75 & 0.25 \\ \hline
 5 & 6 & 4 & 2 & 0.67 & 0.33 \\ \hline
 6 & 6 & 4 & 3 & 0.67 & 0.50 \\ \hline
 7 & 6 & 6 & 2 & 1.0 & 0.33 \\ \hline \hline
 Mean $\pm$ standard  deviation & $6.7\pm1.9$ & $4.9\pm1.1$ & $2.1\pm0.7$ & 0.75 $\pm$ 0.18 & 0.36 $\pm$ 0.21 \\
\end{tabularx}
\caption{Analysis of branching microtubule nucleation efficiency from TPX2 droplets \textit{in vitro}. Data from \cite{Alfaro-Aco2019} was included in this analysis.}
\end{table}

\newpage

\begin{table}[h!] \label{table:s2}
\centering
\begin{tabularx}{\textwidth}{XXXX}
Experimental method & Buffer and reaction components & TPX2 concentration & Microtubule dilution \\ \hline \hline
 TIRFM (TPX2 alone) & CSF-XB with 50-60 mM KCl, 1 mM GTP, and equimolar bovine tubulin & 0.1, 1 $\mu$M & 1:1000-1:2000 \\ \hline
 TIRFM (Branching reconstitution) & BRB80 with 30 mM KCl, 1 mM GTP, 5 mM BME, 0.075\% (w/v) methylcellulose (4000 cP), 1\% (w/v) glucose, 0.02\% (v/v) Brij-35, 250 nM glucose oxidase, 64 nM catalase, 1 mg/mL BSA, 19 $\mu$M unlabeled bovine tubulin and 1 $\mu$M Cy5-labeled bovine tubulin & 0.05 $\mu$M & 1:2000 \\ \hline
 EM (TPX2 alone) & BRB80 & 0.1 $\mu$M & 1:500\\ \hline
 EM (TPX2, augmin, $\gamma$-TuRC) & BRB80 with 50 nM augmin and $\gamma$-TuRC & 0.05 $\mu$M & 1:500 \\ \hline
 AFM & BRB80 & 0.05--1.2 $\mu$M & 1:10\\
\end{tabularx}
\caption{A summary of the \textit{in vitro} conditions used in this study. In one set of conditions used for TIRFM (1 $\mu$M TPX2 and CSF-XB with 50-60 mM KCl, 1 mM GTP, and 1 $\mu$M bovine tubulin), we observed droplets forming within a few minutes (\hyperref[fig:1]{Fig.~1a}). In another set of conditions used for AFM and EM ($\simeq$ 0.05--1.2 $\mu$M TPX2 and BRB80 with no GTP or bovine tubulin), we observed droplets forming over tens of minutes (\hyperref[fig:2]{Fig.~2}). Therefore, different physicochemical conditions affect the timescale of the capillary physics, which is consistent with the hydrodynamic theory. In addition, the film thickness---and therefore the spacing between droplets---is sensitive to both the TPX2 concentration in bulk and the amount of microtubules in the sample (\hyperref[fig:s6]{Fig.~S6b}, \hyperref[theory:1-2]{Supplement 1.2}). Thus, the droplet spacing may vary between experimental methods.}
\end{table}

\newpage

\begin{table}[h!] \label{table:s3}
\centering
\begin{tabularx}{10cm}{XX}
TPX2 concentration $\pm$ estimated range ($\mu$M) & Average peak wavelengths $\pm$ standard deviation (nm) \\ \hline \hline
 0.1 $\pm$ 0.05 & 220 $\pm$ 50\\ \hline
 0.2 $\pm$ 0.1 & 260 $\pm$ 20\\ \hline
 0.6 $\pm$ 0.3 & 300 $\pm$ 30\\ \hline
 0.8 $\pm$ 0.4 & 320 $\pm$ 330 \\
\end{tabularx}
\caption{The peak wavelength for each concentration tested. Peak wavelengths are reported as the peak from bootstrapped average spectra $\pm$ standard deviation of the peak across the bootstrapped distributions. Note that for the highest concentration tested,  0.8 $\pm$ 0.4 $\mu$M, longer line scans were taken using a 4$\times$4 $\mu m$ window. A longer line scan samples a greater number of low frequencies that have high spectral power. This increases the error on our wavelength measurement, which is taken from bootstrapped distributions over all individual spectra.}
\end{table}

\newpage

\begin{table}[h!] \label{table:s4}
\centering
\begin{tabularx}{\textwidth}{XXXXX}
 EM Grid & \# of $\gamma$-TuRCs on grid surface & \# of $\gamma$-TuRCs on bare MTs  & Total \# of $\gamma$-TuRCs & \# of $\gamma$-TuRCs on bare MTs 

/ 

\# of $\gamma$-TuRCs on grid surface \\ \hline \hline
 1 & 40 & 0 & 40 & 0.00 \\ \hline
 2 & 36 & 3 & 39 & 0.08 \\ \hline
 3 & 51 & 4 & 55 & 0.08 \\ \hline \hline
 Mean $\pm$ standard  deviation & $42\pm8$ & $2\pm2$ & $45\pm9$ & $0.05\pm0.05$ \\
\end{tabularx}
\caption{Analysis of $\gamma$-TuRC localization to microtubules (MTs) without TPX2 and augmin using electron microscopy (EM). Results are normalized accounting for total MT length per grid. Data from \cite{Alfaro-Aco2019} was included in this analysis.}
\end{table}

\begin{table}[H] \label{table:s5}
\centering
\begin{tabularx}{\textwidth}{XXXXXXX}
EM Grid & \# of $\gamma$-TuRCs on grid surface & \# of $\gamma$-TuRCs on bare MTs & \# of $\gamma$-TuRCs in droplets on MTs & Total \# of $\gamma$-TuRCs & \# of $\gamma$-TuRCs in droplets on MTs

/ 

\# of $\gamma$-TuRCs on bare MTs & \# of $\gamma$-TuRCs on MTs

/

\# of $\gamma$-TuRCs on grid surface \\ \hline \hline
 1 & 30 & 4 & 10 & 44 & 2.5 & 0.47 \\ \hline
 2 & 34 & 2 & 13 & 49 & 6.5 & 0.44 \\ \hline
 3 & 24 & 4 & 21 & 49 & 5.3 & 0.52 \\ \hline \hline
 Mean $\pm$ standard deviation & $29\pm5$ & $3\pm1$ & $15\pm6$ & $47\pm3$ & $4.8\pm2.0$ & $0.48\pm0.04$ \\
\end{tabularx}
\caption{Analysis of $\gamma$-TuRC localization to microtubules (MTs) in the presence of TPX2 and augmin using electron microscopy (EM). Results are normalized accounting for total MT length per grid. Data from \cite{Alfaro-Aco2019} was included in this analysis.}
\end{table}

\newpage

\section{Supplementary movie captions} \label{movieCaptions}

\href{https://drive.google.com/open?id=1tucttt7pRN6xlIdOjQU1RBxL0GzmhpgC}{\textbf{Supplementary Movie 1.}} Rayleigh-Plateau instability of TPX2 on microtubules visualized using TIRF microscopy. $1~\mu\mathrm{M}$ GFP-TPX2 was spiked onto a passivated glass surface covered with Alexa568-labeled microtubules at $t=0~\mathrm{s}$. TPX2 coats the microtubule and then proceeds to break up into droplets. Scale bar is $1~\mu\mathrm{m}$. 

\noindent \href{https://drive.google.com/file/d/1JcGqQLrMGLoQ5b-1vkoJ9fuISq-UXJBw/view?usp=sharing}{\textbf{Supplementary Movie 2.}} Microtubule branches nucleating from TPX2 droplets on a preexisting microtubule. During acquisition, only the soluble tubulin channel was imaged to enable capturing nucleation and polymerization of branched microtubules at high temporal resolution. Scale bar is $5~\mu\mathrm{m}$.

\noindent \href{https://drive.google.com/open?id=1cS-YTCxQVYJkrR9x-x-Ul8bZ3CtQh58A}{\textbf{Supplementary Movie 3.}} Branching microtubule nucleation visualized using TIRF microscopy. $0.1~\mu\mathrm{M}$ TPX2 is added to \textit{X. laevis} meiotic cytosol purified from eggs. TPX2 coats the mother microtubule from which daughter microtubules then nucleate, leading to an autocatalytic branched network. Frame dimensions are $16~\mu\mathrm{m} \times 22~\mu\mathrm{m}$. 

\noindent \href{https://drive.google.com/open?id=1U3z1bUpFgl_GClE1fUYhAKzR9uJHuzkv}{\textbf{Supplementary Movie 4.}} Rayleigh-Plateau instability of TPX2 on microtubules probed using AFM. $0.2\pm 0.1~\mu\mathrm{M}$ GFP-TPX2 was spiked onto a mica surface covered with microtubules during acquisition. The uniform film of TPX2 is established at $t=0~\mathrm{s}$, after which the film breaks up into droplets. Frame dimensions are $2~\mu\mathrm{m} \times 2~\mu\mathrm{m}$.

\newpage

\bibliography{bib_rayleigh_plateau_on_MTs.bib}
\bibliographystyle{naturemag}

\end{document}